\documentclass[]{article}
\usepackage{graphicx}
\usepackage{epstopdf}
\usepackage{amsmath}
\usepackage{amssymb}
\usepackage{textcomp}
\usepackage{verbatim}
\usepackage{cite}
\everymath{\displaystyle}
\usepackage[aboveskip=5pt,belowskip=10pt]{caption}
\usepackage[aboveskip=1pt]{subcaption}
\title{Stable-regime probability in argumentary Duffing oscillators}
\author
{
Daniel Cintra (corresponding author)\\
Universit\'{e} Paris-Est, Laboratoire Navier (UMR 8205),\\
CNRS, ENPC, IFSTTAR,\\
F-77455 Marne La Vall\'{e}e, Cedex 2, France\\
email: daniel.cintra@enpc.fr\\
and\\
Pierre Argoul \\
IFSTTAR, Laboratoire MAST-SDOA,\\
F-77455 Marne La Vall\'{e}e, Cedex 2, France\\
email: pierre.argoul@ifsttar.fr
}
\date{}
\everymath{\displaystyle}
\providecommand{\keywords}[1]{\textbf{\textit{Keywords---}} #1}
\title{Nonlinear argumental oscillators: Stability criterion and attractor's capture probability}
\begin{document}
\maketitle
\begin{abstract}
The behaviour of a space-modulated, so-called ''argumental'' oscillator is studied, which is represented by a model having an even-parity space-modulating function. Analytic expressions of a stability criterion and of discrete energy levels are given. Using an integrating factor and a Van der Pol representation in the (amplitude, phase) space, an approximate implicit closed-form of the solution is given. The probability to enter a stable-oscillation regime from given initial conditions is calculated in symbolic form. These results allow an analytic approach to stability and bifurcations of the system. They also allow an assessment of the risk of occurrence of sustained large-amplitude oscillations, when the phenomenon is to be avoided, and an assessment of the conditions to apply to obtain oscillations whenever the phenomenon is desired.
\end{abstract}
\newpage
\tableofcontents
\newpage
\section{Introduction}
In the 1920s, physicists were searching for a device to divide the mains current frequency in order to manufacture mains-driven clocks. As no electronics were available, they studied various inherently frequency-dividing oscillators. Among them was a pendulum designed by B\'ethenod in 1929 \cite{Bethenod_38_AS}, that oscillated at a low frequency, typically 1~Hz, when driven by the mains at 50~Hz. This design was inspired by a remark made by Soulier \cite{Soulier_25} in 1928 about an oscillating bar. Before, Cornu \cite{Cornu} and F\'ery \cite{Fery} had designed pendulums based on magnetic forces, aimed at the synchronization of clocks, but with a pendulum frequency equal to the excitation frequency. B\'ethenod's oscillator was a pendulum fitted with a steel sphere at the tip of the rod. The sphere could sense the external electromagnetic force only when it was near the lower equilibrium position of the pendulum. Thus, there was a spatial modulation of the force. The force was due to a magnetic field created by a solenoid with vertical axis, carrying an alternating current. The force could only be attractive. B\'ethenod presented observations and calculation using a perturbation method. He did not go deeper into this phenomenon. In this paper, this type of oscillator is referred to as ``B\'ethenod's pendulum'' or ``Type A oscillator'' \cite{Doubochinski_Book}.\\

In the 1960s, Russian researchers studied an oscillator subjected to a spatially-localized external force; the oscillator operated at a frequency much lower than that of the external force. D.~I.~Penner et al. coined the term ``argumental oscillations'' \cite{Penner_Doubo_72,Penner_Doubo_73} from the fact that the interaction between the oscillator and the excitation depends on the ''argument'' of a space-localization function, which we call the $H$-function hereinafter. The oscillator was a pendulum fitted with a permanent electric charge at the tip of the rod. The charge crossed a parallel-plate capacitor connected to an harmonic voltage, and could sense a force only when passing through the capacitor. From that time on, Doubochinski \cite{Doubochinski_Book}, who was in Penner's team, studied this type of oscillator, as well as other related types, and made many publications. He used a pendulum fitted with a permanent magnet at the tip of the rod, whose magnetic moment was aligned with the rod. In this set-up, the magnet can sense the external electromagnetic force only when it is near the lower equilibrium position. The force is due to a magnetic field created by a coil with horizontal axis perpendicular to the plane of displacement of the rod and the tip. This force can be attractive or repulsive, depending on the polarity of the current producing the magnetic field. In this paper, this type of oscillator is referred to as ``Doubochinski's pendulum'' or ``Type B oscillator''. Doubochinski modelled this phenomenon and produced mathematical results regarding the resolution of the system \cite{Doubo_EDF_91}. He modelled the spatial localization of the interaction zone by means of a gate function. He also designed physical models and carried out a vast number of measurements. He also built many devices using this phenomenon.\\

Motion of oscillators at a frequency lower than the excitation frequency has been studied: argumental oscillations, observed in ~\cite{Treilhou_00}. An electronic argumental oscillator with a gate function used as dependent-variable localization function has been studied in \cite{Cretin_Vernier}. A new formula about the magnetic interaction between an external force and an argumental oscillator is given in \cite{Cintra_Argoul_Dynolin_2013}. Preliminary experimental results about six argumental oscillators are given in \cite{Cintra_Argoul_Dynolin_2014}.
Modeling and experimental results about six argumental oscillators are given in \cite{Cintra_Argoul_JVC_2016}.
A symbolic formula giving the stable-regime establishment probability of an argumental oscillator is exposed in \cite{Cintra_Argoul_Vishno_2014}.

\section{Canonical second-order equation of motion.}
To simplify the expression of the system behaviour, one classically uses the reduced time $\tau=\omega_0 t$, where $\omega_0$ is the natural angular velocity of the oscillator. Using from now on the dot notation to refer to the derivatives with respect to $\tau$, we shall distinguish two types of oscillator, which we call ``Type A'' and ``Type B''. The second-order equation of motion for the Type A oscillators is:
\begin{equation}
\ddot{\alpha}+2 \beta \dot{\alpha}+\alpha +\mu \alpha^3=A H(\alpha) \sin^2\left(\frac{\nu}{\omega_0} \tau\right)
\label{eq:Equ_motion_Bethenod_reduced_time}
\end{equation}
where $\beta$ is a dissipation coefficient, $\mu$ is the Duffing coefficient, $A$ is a constant, $H$ is an odd function of $\alpha$, and $\nu$ is the external harmonic excitation's angular velocity. An example of a Type-A $H$-function is
\begin{equation}
H(\alpha)=\frac{\alpha}{(1+\gamma \alpha^2)^3},
\label{eq:Type_A_H_function}
\end{equation}
with $\gamma$ being a constant.\\
For the Type B oscillators, the equation is:
\begin{equation}
\ddot{\alpha}+2 \beta \dot{\alpha}+\alpha +\mu \alpha^3=A H(\alpha) \sin\left(\frac{\nu}{\omega_0} \tau\right)
\label{eq:Equ_movt_Doubochinski_reduced_time}
\end{equation}
where H is an even function of $\alpha$. An example of a Type-B $H$-function is
\begin{equation}
H(\alpha)=\frac{1-\gamma \alpha^2}{(1+\gamma \alpha^2)^2},
\label{eq:Type_B_H_function}
\end{equation}
with $\gamma$ being a constant.\\
The ``Type A'' oscillator in this paper is a Type II-1 oscillator as of our article \cite{Cintra_Argoul_JVC_2016}, while the ``Type B'' oscillator is a Type I-2 oscillator.\\
We can remark that the equations \eqref{eq:Equ_motion_Bethenod_reduced_time} and \eqref{eq:Equ_movt_Doubochinski_reduced_time} are similar. The only formal difference is in the expression of the external force, which is in $\sin^2\left(\frac{\nu}{\omega_0}\tau\right)$ for the Type A oscillator, and in $\sin\left(\frac{\nu}{\omega_0}\tau\right)$ for the Type B oscillator. This will lead to parity considerations in the averaging process, but we shall see later on that this doesn't imply a formal difference in the averaged system.\\
\cite{Doubo_EDF_91} used a coarser approximation to a Type-B $H$-function by putting $H(\alpha)=rect(2 \frac{\alpha}{\alpha_0})$, where rect denotes the rectangular function, i.e. $H(\alpha)=1$ if $\lvert \alpha \rvert <\alpha_0$ and $H(\alpha)=0$ otherwise. We denote this model by the Type-C model. This approximation was sufficient to elaborate averaged equations and to derive an expression of the amplitude of the external force as a function of the oscillator amplitude, leading to an explanation of a discrete set of stable amplitudes. We shall use herein our smoother and more precise Type-B $H$-function, as of \eqref{eq:Type_B_H_function}, with the advantage of handling a $H$-function which is $C^\infty$: this will allow us to eliminate artifacts in the ''$A$-function'' of the Type-C model, and to derive an approximate symbolic solution leading to an expression of the capture probabilities.
\section{Calculus workflow.}
Having available the reduced-time second-order differential equation of motion for both oscillators, one considers that a perturbation method could be an appropriate approach, because the oscillator is almost always in a free-run mode. Only at certain narrow locations in space will it ``feel'' the external force. Moreover, this force is of small amplitude. Keeping the expressions under symbolic form, we shall go through four steps to get to the capture probabilities.\\
The first and second steps are classical, and have been described by Poincar\'e xxx and \cite{Bogolioubov_Book}, and used by \cite{Doubo_EDF_91}. So we shall only outline the calculus for these two steps. Our contribution to the two first steps is the symbolic expression of the Fourier series for the $H$-functions of the Type-A and Type-B oscillators and the Van der Pol polar representation of the averaged amplitude and phase.
\begin{itemize}
\item The first step of the calculus is to replace the second-order differential equation of motion by two first-order equations to get the classical standard system of equations.
\item The second step is to use the averaging method to obtain an averaged system of equations. The idea here is to use a Fourier series of the $H$-function to apply the averaging calculus. As the external force is harmonic, we can expect simplifications.
\end{itemize}
Our contribution consists of the third and fourth steps, which are as follows, and which will be detailed hereinafter:
\begin{itemize}
\item The third step is to find an integrating factor to approximately solve the averaged system, while keeping the symbolic form of the equations.
\item The fourth step is to use the approximate symbolic solution to derive the capture probabilities, i.e. the probabilities to enter a stable regime.
\end{itemize}
\section{First step: building the standard system of equations.}
Starting from the equation of motion under its general form \eqref{eq:Equ_movt_Doubochinski_reduced_time}, one defines a function X by:\\
\begin{equation}
X(\tau)=-2 \beta \frac{d\alpha}{d\tau}-\mu \alpha^3+A H(\alpha) \sin\left(\frac{\nu}{\omega_0} \tau\right).
\label{eq:eq2_2}
\end{equation}
Thus equation \eqref{eq:Equ_movt_Doubochinski_reduced_time} can be rewritten:
\begin{equation}
\frac{d^2\alpha}{d\tau^2}+\alpha=X(\tau, \alpha, \dot{\alpha}).
\label{eq:eq2_4}
\end{equation}
By observing the experimental oscillators and corresponding numerical simulations, one concludes that the motion is close to that of a free-running oscillator, with slowly varying amplitude and phase. Hence one introduces the slow-varying amplitude $a(\tau)$ and phase $\varphi(\tau)$ as two new independent variables, which will replace the variables $\alpha$ and $\dot{\alpha}$. The motion expressed as a function of t will be $\alpha(t)=a(t) \sin(\omega t+\varphi(t))=a(\tau) \sin(\rho \tau+\varphi(\tau))$, where $\omega$ is a parameter close to $\omega_0$, and $\rho=\frac{\omega}{\omega_0}$. As these two new independent variables are chosen, we found it natural to introduce a Van der Pol representation, with $a$ as absciss$a$ and $\varphi$ as ordinates. Alternatively, we shall also use a polar Van der Pol representation, i.e. $a$ as radius and $\varphi$ as angle.

Define the change of variables by putting:
\begin{align}
\alpha(\tau)=a(\tau) \sin(\rho \tau+\varphi(\tau)),
\label{eq:eq2_5}\\
\dot{\alpha} (\tau)=a(\tau) \rho \cos(\rho \tau+\varphi(\tau)).
\label{eq:eq2_6}
\end{align}
This is natural, because equation \eqref{eq:eq2_6} is obtained by derivating equation \eqref{eq:eq2_5} with $a$ and $\varphi$ taken as constant. This is simply the implementation of the physical observation that $a$ and $\varphi$ vary slowly with respect with the period of the free-running oscillator.\\
Differentiating \eqref{eq:eq2_5} and comparing the result with \eqref{eq:eq2_6}, one gets:
\begin{equation}
\dot{a} \sin(\rho \tau+\varphi)+a \dot{\varphi} \cos(\rho \tau+\varphi)=0
\label{eq:eq2_8}
\end{equation}
Differentiating \eqref{eq:eq2_6} and putting the result into \eqref{eq:eq2_4}, one gets:
\begin{equation}
\dot{a}=\frac{\cos(\theta)}{\rho} (X(\tau, a \sin(\theta), a \rho \cos(\theta))+a \sin(\theta) (\rho^2-1)),
\label{eq:eq2_13}
\end{equation}
where $\theta=\rho \tau+\varphi$.
This is the first differential equation involving only the two new variables $a$ and $\varphi$.\\
Substituting in \eqref{eq:eq2_8} the expression \eqref{eq:eq2_13} for $\dot{a}$, one gets:
\begin{equation}
\dot{\varphi}=-\frac{1}{a \rho} (a \sin(\theta) (\rho^2-1)+X(\tau, a \sin(\theta), a \rho \cos(\theta))) \sin(\theta)
\label{eq:eq2_14}
\end{equation}
which is the second differential equation involving only the two new variables $a$ and $\varphi$.
\section{Second step: averaging the standard system of equations.}
\subsection{Case H = even function (Type B oscillators).}
In this section, one uses the variable $\theta=\rho \tau+\varphi$ as defined in the previous section, and one supposes that $H(\alpha)$ is an even function of $\alpha$. That is, we focus on oscillators of Type B. We shall study the oscillators of Type A thereafter.\\
One has to average equations \eqref{eq:eq2_13} and \eqref{eq:eq2_14} with respect to $\theta$. To do so, one forms the Fourier series of $H(\alpha)=H(a \sin(\theta))$.
\subsubsection{Averaging equation \eqref{eq:eq2_13}.}
In equation \eqref{eq:eq2_13}, replacing $X(\tau, a \sin(\theta), a \rho \cos(\theta))$ by its expression given in \eqref{eq:eq2_2}, one gets, taking into account \eqref{eq:eq2_5} and \eqref{eq:eq2_6}:
\begin{align}
\dot{a}&= \frac{\cos(\theta)}{\rho} \left(-2 \beta \dot{\alpha}-\mu \alpha^3+A H(\alpha) \sin\left(\frac{\nu}{\omega_0} \tau\right)+a \sin(\theta) (\rho^2-1)\right) \nonumber\\
 &= \frac{\cos(\theta)}{\rho} \left(-2 \beta a \rho \cos(\theta)-\mu a^3 \sin^3(\theta)+A H(a \sin(\theta)) \sin\left(\frac{\nu}{\omega_0} \tau\right)+a \sin(\theta) (\rho^2-1)\right). \nonumber
\end{align}
One has to average $\dot{a}$ with respect to $\theta$.\\
Knowing that $\overline{\cos^2(\theta)}=\frac{1}{2}$, \hspace{1pt} $\overline{\cos(\theta) \sin^3(\theta)}=0$,\hspace{1pt} and $\overline{\cos(\theta) \sin(\theta)}=0$, one gets :
\begin{equation}
\overline{\dot{a}} = -\beta a+\frac{A}{\rho} \overline{H(a \sin(\theta))\cos(\theta) \sin\left(\frac{\nu}{\omega_0} \tau\right)}
\end{equation}
Knowing that H is an even function of $\alpha$, one introduces the Fourier series of $H(a \sin(\theta))$, namely
\begin{equation}
H(a \sin(\theta))=\sum_{q=0}^{+\infty}c_q(a) \cos(2 q \theta)
\end{equation}
with
\begin{equation}
c_q(a)=\frac{2}{\pi} \int_{0}^{\pi} H(a \sin(\eta)) \cos(2 q \eta) d\eta.
\end{equation}
At this point, define a real number n by $n=\frac{\nu}{\omega}=\frac{\nu}{\rho \omega_0}$. It holds:
\begin{align}
\overline{H(a \sin(\theta))\cos(\theta) \sin\left(\frac{\nu}{\omega_0} \tau\right)}&= \overline{\sum_{q=0}^{+\infty}c_q(a) \cos(2 q \theta) \cos(\theta) \sin\left(\frac{\nu}{\omega_0} \tau\right)} \nonumber\\
 & =\sum_{q=0}^{+\infty}c_q(a) \overline{\cos(2 q \theta) \cos(\theta) \sin\left(\frac{\nu}{\omega_0} \tau\right)} \nonumber \\
 & =\begin{cases}
 -\frac{1}{4} \sin(n \varphi) (c_m(a)+c_p(a)) & \text{if n is an odd integer,} \nonumber\\
 0 & \text{otherwise,}
 \end{cases}
\end{align}
with $m=\frac{n-1}{2}$ and $p=\frac{n+1}{2}$.\\
Consequently, if n is an odd integer, one gets the first averaged equation :
\begin{equation}
\dot{a} = -\frac{A}{4 \rho} S(a) \sin(n \varphi)-\beta a,
\label{eq:eq2_24}
\end{equation}
with $m=\frac{n-1}{2}$, $p=\frac{n+1}{2}$, and $S(a)=c_m(a)+c_p(a)$.\\
and otherwise, one has:
\begin{equation}
\overline{\dot{a}} = -\beta a.
\label{eq:1st_averaged_eqn_n_not_odd}
\end{equation}
That is, if n is not an odd integer, the averaged equation is the same as for n odd integer, except that one must make $A=0$.\\
The various possible values of n will be discussed hereafter.
\subsubsection{Averaging equation \eqref{eq:eq2_14}.}
From equation \eqref{eq:eq2_14}, one gets, in the same way:
\begin{align}
\dot{\varphi} & =  -\frac{1}{a \rho} (a \sin(\theta) (\rho^2-1)-2 \beta \dot{\alpha}-\mu \alpha^3+A H(\alpha) \sin\left(\frac{\nu}{\omega_0} \tau\right)) \sin(\theta) \nonumber\\
       &= \frac{\sin(\theta)}{a \rho} \left(-a \sin(\theta) (\rho^2-1)+2 \beta a \rho \cos(\theta)+\mu a^3 \sin^3(\theta)-A H(a \sin(\theta)) \sin\left(\frac{\nu}{\omega_0} \tau\right)\right). \nonumber
\end{align}
One has to average $\dot{\varphi}$ with respect to $\theta$.
In the same way as for the averaging of equation \eqref{eq:eq2_13}, one has:
\begin{align}
\overline{H(a \sin(\theta))\sin(\theta) \sin\left(\frac{\nu}{\omega_0} \tau\right)}=
 \begin{cases}
 \frac{1}{4} (c_m(a)+c_p(a)) \cos(n \varphi) & \text{if n is an odd integer,} \nonumber\\
 0 & \text{otherwise}.
 \end{cases}
\end{align}
Knowing that $\overline{\sin^2(\theta)}=\frac{1}{2}$, \hspace{1pt} $\overline{\sin^4(\theta)}=\frac{3 \pi}{8}$,\hspace{1pt} and $\overline{\cos(\theta) \sin(\theta)}=0$, one gets, if n is an odd integer:
\begin{equation}
\overline{\dot{\varphi}}=\frac{1-\rho^2}{2 \rho}+\frac{3 \pi}{8} \frac{\mu a^2}{\rho}-\frac{A}{4 a \rho} \cos(n \varphi) (c_m(a)-c_p(a)),
\label{eq:eq2_17}
\end{equation}
and otherwise:
\begin{equation}
\overline{\dot{\varphi}}=\frac{1-\rho^2}{2 \rho}+\frac{3 \pi}{8} \frac{\mu a^2}{\rho}.
\end{equation}
That is, here again, as for the averaging of equation \eqref{eq:eq2_13}, if n is not an odd integer, the averaged equation is the same as for n odd integer, except that one must make $A=0$.\\
In equation \eqref{eq:eq2_17}, define $a_n$ by:
\begin{equation}
\rho^2=\left(\frac{1}{n} \frac{\nu}{\omega_0}\right)^2=\frac{3 \mu a_n^2}{4}+1.
\label{eq:definition_of_a_n}
\end{equation}
Then $\frac{1-\rho^2}{2 \rho}=-\frac{3 \mu a_n^2}{8}$, and equation \eqref{eq:eq2_17} can be written as follows:
\begin{equation}
\dot{\varphi}=\frac{3\mu}{8 \rho}(a^2-a_n^2)-\frac{A}{4 a \rho} \cos(n \varphi) D(a).
\label{eq:eq2_19}
\end{equation}
with $m=\frac{n-1}{2}$, $p=\frac{n+1}{2}$, and $D(a)=c_m(a)-c_p(a)$.\\
And if n is not an odd integer, one has, putting $A=0$ in \eqref{eq:eq2_19}:
\begin{equation}
\dot{\varphi}=\frac{3\mu}{8 \rho}(a^2-a_n^2).
\label{eq:2nd_averaged_eqn_n_not_odd}
\end{equation}
\subsubsection{Conclusion: the averaged system for Type B oscillators.}
The averaged system can now be formed, consisting of equations \eqref{eq:eq2_24} and \eqref{eq:eq2_19}:
\begin{align}
 \begin{cases}
  \dot{a} &= -\frac{A}{4 \rho_n} S(a) \sin(n \varphi)-\beta a \\
  \dot{\varphi}&= \frac{3\mu}{8 \rho_n}(a^2-a_n^2)-\frac{A}{4 a \rho_n} \cos(n \varphi) D(a).
  \label{eq:averaged_system}
 \end{cases}
\end{align}
\paragraph{The possible values of n. Critical value of n.}~\\
In this paragraph, we shall study the link between $\nu$, $\omega_0$, n, and the amplitudes of the free oscillator. This will enable us to give a physical sense to $\rho$.\\
Recall the system of equations \eqref{eq:1st_averaged_eqn_n_not_odd}, \eqref{eq:2nd_averaged_eqn_n_not_odd}, where the external force is null ($A=0$):
\begin{align}
 \begin{cases}
\dot{a} &= -\beta a\\
\dot{\varphi}&=\frac{3\mu}{8 \rho}(a^2-a_n^2).
 \end{cases}
 \label{eq:system_when_A_0}
\end{align}
Now consider the case of the free undamped oscillator, looking for the stable regime condition. We have $\beta=0$ and therefore, the system \eqref{eq:system_when_A_0} resolves to $a=a_0$ and $\dot{\varphi}=\frac{3\mu}{8 \rho}(a_0^2-a_n^2)$, where $a_0$ is the initial value of a. The regime will be stable if $\dot{\varphi}$ is null, that is, if $a_0=a_n$. Thus, the value $a_n$ which appears in equation \eqref{eq:definition_of_a_n} is the amplitude of the free undamped oscillator. This equation shows that we must distinguish between the case $\mu >0$ and the case $\mu <0$. Let's introduce an integer critical value for n: $n_{crit}=\left[ \frac{\nu}{\omega_0}\right]$, where the square bracket notation means here: ``integer part of''. Define a real $\varepsilon$ by $\frac{\nu}{\omega_0}=\left[ \frac{\nu}{\omega_0}\right]+\varepsilon$; we have $0 \leqslant \varepsilon<1$. From equation \eqref{eq:definition_of_a_n}, we get $a_n^2=\frac{4}{3 \mu} \left(\left(\frac{1}{n} \frac{\nu}{\omega_0}\right)^2-1\right)$. Thus if $\mu$ is negative, we must have $\left(\frac{1}{n} \frac{\nu}{\omega_0}\right)^2-1<0$, i.e. $n>\frac{\nu}{\omega_0}=n_{crit}+\varepsilon$, which reduces to $n\geqslant n_{crit}+1$. We can take n as big as we want, provided we stay in the validity domain of the averaging method. The energy of the oscillator increases as n increases.\\
Conversely, if $\mu$ is positive, we must have $n<\frac{\nu}{\omega_0}=n_{crit}+\varepsilon$, which reduces to $n\leqslant n_{crit}$; we must also have $n\geqslant 3$, because n is odd and must be greater than 1 (otherwise, we have a classical forced-oscillations system). We have a finite number of possible stable amplitudes, and the energy increases as n decreases.\\
From equation \eqref{eq:definition_of_a_n}, we can get the physical sense of $\rho$: this value gives an indication of the energy of the system, and takes a discrete set of values. This is why from now on, we shall write $\rho_n$ instead of $\rho$, because to each value of n is associated a value of $\rho$.\\
This discrete series of stable amplitudes and energy levels has been discussed in \cite{Doubo_EDF_91}.
\subsection{Case H = odd function (Type A oscillators).}
In this section, we suppose that $H(\alpha)$ is an odd function of $\alpha$. That is, we focus on oscillators of Type A.\\
The main difference between those two types of oscillator lies in the Fourier series of the function $H(a \sin(\theta))$. Because this function of $\theta$ is odd, one has:
\begin{equation}
H(a \sin(\theta))=\sum_{q=0}^{+\infty}c_q(a) \sin((2 q+1) \theta)
\end{equation}
with
\begin{equation}
c_q(a)=\frac{1}{\pi} \int_{0}^{2 \pi} H(a \sin(\eta)) \sin((2 q+1) \eta) d\eta,
\end{equation}
and therefore:
\begin{align}
\overline{H(a \sin(\theta))\cos(\theta) \sin^2(\frac{\nu}{\omega_0} \tau)}=
 \begin{cases}
 -\frac{1}{8} (c_m(a)+c_p(a)) \sin(n \varphi) & \text{if n is an even integer,} \nonumber\\
 0 & \text{otherwise,}
 \end{cases}
\end{align}
with $n=\frac{2 \nu}{\omega}=\frac{2 \nu}{\rho_n \omega_0}$, $m=\frac{n}{2}-1$ and $p=\frac{n}{2}$. Hence, in the same way as for the Type B oscillator, one gets the following system of two averaged equations:\\
If $n=\frac{2 \nu}{\omega}=\frac{2 \nu}{\rho_n \omega_0}$ is an even integer:
\begin{align}
 \begin{cases}
  \dot{a} &= -\frac{A}{8 \rho_n} S(a) \sin(n \varphi)-\beta a \\
  \dot{\varphi}&= \frac{3\mu}{8 \rho_n}(a^2-a_n^2)-\frac{A}{8 a \rho_n} \cos(n \varphi) D(a),
  \label{eq:eq2_23}
 \end{cases}
\end{align}
with $S(a)=c_m(a)+c_p(a)$, $D(a)=c_m(a)-c_p(a)$, $m=\frac{n}{2}-1$ and $p=\frac{n}{2}$.\\
If n is not an even integer:
\begin{align}
 \begin{cases}
  \dot{a} & = -\beta a \\ 
  \dot{\varphi} & = \frac{3\mu}{8 \rho_n}(a^2-a_n^2).
  \label{eq:eq2_21}
 \end{cases}
\end{align}
We can see that this averaged system is formally similar to the averaged system \eqref{eq:averaged_system} of the Type B oscillator, in which we would substitute $A/2$ for $A$. The difference is in the expression of the functions S(a) and D(a). Therefore, we shall use from now on the averaged system of the Type B oscillator whenever we shall have to discuss a point which is not specific to the $H$-function, i.e. whenever we will not have to give an explicit expression for the functions $S(a)$ and $D(a)$.
\subsection{Symbolic expressions of the $S(a)$ and $D(a)$ functions.}
In order to be able to draw plots for given examples, we have to know the explicit expressions of the $S$ and $D$ functions for a given oscillator. Therefore, we shall calculate the $c_q(a)$ coefficients of the $H$-function for one Type-A oscillator and one Type-B oscillator. From these coefficients, we will be able to express the $S$ and $D$ functions. We shall also calculate these coefficients for the Type-C oscillator, to show the artifacts it introduces.
\subsubsection{Type A.}
As an example of Type A oscillator, we take the original B\'ethenod pendulum and our $H$-function as of \eqref{eq:Type_A_H_function}.
We have to calculate the generic r-th term of the Fourier series of $H(a sin(\theta)$, namely:
\begin{align}
c_r(a)&=\frac{1}{\pi} \int_{0}^{2 \pi} H(a \sin(\eta)) \sin((2 r+1) \eta) d\eta \nonumber \\
&= \frac{1}{\pi} \int_{0}^{2 \pi} \frac{a \sin(\eta)}{(1+\gamma a^2 \sin^2(\eta))^3} \sin((2 r+1) \eta) d\eta
\end{align}
We define $L(b,q,m)$ as:
\begin{equation}
L(b,q,m)=\int_0^{2 \pi} \frac{sin(\eta) \sin(q \eta)}{(1+b \sin^2(\eta))^m} d\eta. \nonumber
\end{equation}
Hence we have:
\begin{equation}
c_r=\frac{a}{\pi} L(\gamma a^2,2 r+1,3). \nonumber
\end{equation}
and
\begin{align}
S=c_{n-1}+c_{n+1}=\frac{a}{\pi} (L(\gamma a^2,2 n-1,3)+L(\gamma a^2,2 n+3,3)) \nonumber \\
D=c_{n-1}-c_{n+1}=\frac{a}{\pi} (L(\gamma a^2,2 n-1,3)-L(\gamma a^2,2 n+3,3)) \nonumber
\end{align}
Then, using \cite[\S 3.616-7]{Gradshteyn_07}, we find:
\begin{align}
L(b,q,m)&=\frac{1}{2} \frac{1}{\left(1+\frac{b}{2}\right)^m} (M(q-1,g,m)-M(q+1,g,m)), \text{with} \nonumber\\
M(s,g,m)&=\int_0^{2 \pi} \frac{cos(s \eta)}{1-g \cos(2 \eta))^m} d\eta \nonumber\\
&=\frac{2^{m+1}}{g^m} \frac{\pi}{f^{n-m} (f^2-1)^{2 m-1}} \sum_{k=0}^{m-1} \binom{m+n-1}{k} \binom{2 m-k-2}{m-1} (f^2-1)^k \nonumber
\end{align}
$\text{with } n=\frac{s}{2}, g=\frac{b}{2+b} \text{ and } f=\frac{1}{g}+\sqrt{\frac{1}{g^2}-1}.$\\
Finally, we find that some approximations are possible. If $q\gg 5$ and $\sqrt{1+b}\gg 1$, we have:
\begin{equation}
L(b,q,3)\approx \frac{\pi}{8 \sqrt{b}} \left(6+4 \frac{q}{b} +36 \frac{q^2}{b^2}\left(1+\frac{106}{36 b}+\frac{308}{36 b^2}+\frac{32}{36 b^3} \right) \right). \nonumber
\end{equation}
And if $q<b$ and $b\gg\frac{106}{36}$, we have:
\begin{equation}
L(b,q,3)\approx \frac{\pi}{8 \sqrt{b}} \left(6+4 \frac{q}{b} +36 \frac{q^2}{b^2} \right). \nonumber
\end{equation}
\subsubsection{Type B.}
As of \eqref{eq:Type_B_H_function}, we have $H(\alpha)=\frac{1-\gamma \alpha^2}{(1+\gamma \alpha^2)^2}$ and $H(a \sin(\eta))=\sum_0^{+\infty}c_q \cos(2 q \theta)$, and therefore:
\begin{equation}
c_q=\frac{1}{\pi} \int_0^{2 \pi} \frac{1-\gamma a^2 \sin^2(\eta)}{(1+\gamma a^2 \sin^2(\eta))^2} \cos(2 q \eta)d\eta, \nonumber
\end{equation}
and we find, using \cite[\S 3.613]{Gradshteyn_07} and after a few manipulations:
\begin{equation}
c_q(a)=2 \: \frac{1+2 q \sqrt{1+\gamma a^2}}{(1+\gamma a^2)^{3/2}} \left(\frac{1-\sqrt{1+\gamma a^2}}{a \sqrt{\gamma}}\right)^{2 q}.
\end{equation}
Hence, with $S=c_{\frac{n-1}{2}}+c_{\frac{n+1}{2}}$ and $D=c_{\frac{n-1}{2}}-c_{\frac{n+1}{2}}$:
\begin{align}
S(a)&=\frac{4 n}{a \sqrt{\gamma} \sqrt{1+\gamma a^2}} \left(\frac{\sqrt{1+\gamma a^2}-1}{a \sqrt{\gamma}}\right)^{n}\\
D(a)&=\frac{4}{a \sqrt{\gamma} (1+\gamma a^2)^{3/2}} \left(\frac{\sqrt{1+\gamma a^2}-1}{a \sqrt{\gamma}}\right)^n (n \sqrt{1+\gamma a^2}-\gamma a^2)\\
\frac{D(a)}{S(a)}&=\frac{1}{\sqrt{1+\gamma a^2}}-\frac{1}{n} \frac{\gamma a^2}{1+\gamma a^2}.
\label{eq:S_D_and_S_D}
\end{align}
\subsubsection{Type C.}
We call Type C the original model from \cite{Doubo_EDF_91}. Taking $H(x)=rect(2 \frac{x}{h})$ and putting $\alpha=\arcsin\left( \frac{h}{a}\right)$ if $a \ge h$ and $\alpha=\frac{\pi}{2}$ if $a < h$, we find $c_q(a)=\frac{4}{\pi q} \sin\left(q \alpha \right)$. It follows that, with $S=c_{\frac{n-1}{2}}+c_{\frac{n+1}{2}}$ and $D=c_{\frac{n-1}{2}}-c_{\frac{n+1}{2}}$:
\begin{align}
S(\alpha)&=\frac{4}{\pi} \left(\frac{\sin\left(\frac{n-1}{2} \alpha\right)}{\frac{n-1}{2}}+\frac{\sin\left(\frac{n+1}{2} \alpha\right)}{\frac{n+1}{2}}\right)\\
D(\alpha)&=\frac{4}{\pi} \left(\frac{\sin\left(\frac{n-1}{2} \alpha\right)}{\frac{n-1}{2}}-\frac{\sin\left(\frac{n+1}{2} \alpha\right)}{\frac{n+1}{2}} \right)\\
\frac{D(\alpha)}{S(\alpha}&=\frac{\tan(\frac{n \alpha}{2})-n \tan(\frac{\alpha}{2})}{n \tan(\frac{n \alpha}{2})-\tan(\frac{ \alpha}{2})}
\label{eq:equ_S_24}
\end{align}
\subsection{Equilibrium.}
We shall discuss herein the averaged system of Type B.
\subsubsection{Stability condition.}
Recall the averaged system of Type B, given in \eqref{eq:averaged_system}:
\begin{align}
\begin{cases}
\dot{a} = -\frac{A}{4 \rho_n} S(a) \sin(n \varphi)-\beta a  \\
\dot{\varphi}=\frac{3\mu}{8 \rho_n}(a^2-a_n^2)-\frac{A}{4 a \rho_n} \cos(n \varphi) D(a)
\end{cases}
\end{align}
We shall write this system in a more general fashion, as follows:
\begin{align}
\begin{cases}
\dot{a} =A f(a) \sin(n \varphi)+g(a)=F(a,\varphi) \\
\dot{\varphi}=A j(a) \cos(n \varphi)+h(a)=G(a,\varphi)
\end{cases}
\label{eq:equil_general}
\end{align}
Let's write the equilibrium condition for this averaged system. Call $a_S$ and $\varphi_S$ the values of $a$ and $\varphi$ at an equilibrium point. Putting $\dot{a}=0$ and $\dot{\varphi}=0$ for $a=a_S$ and $\varphi=\varphi_S$, we get:
\begin{align}
\begin{cases}
A f(a_S) \sin(n \varphi_S)+g(a_S)=0 \\
A j(a_S) \cos(n \varphi_S)+h(a_S)=0
\label{eq:equil_general_2}
\end{cases}
\end{align}
Hence:
\begin{align}
\begin{cases}
A^2 = \frac{g^2(a_S)}{f^2(a_S)}+\frac{h^2(a_S)}{j^2(a_S)}\\[3.0ex]
tan(n \varphi_S) = \frac{g(a_S) j(a_S)}{f(a_S) h(a_S)}
\label{eq:A2_fn_of_aS_and_phiS}
\end{cases}
\end{align}
For instance, for the system of Type B, we get:
\begin{align}
\begin{cases}
A = \frac{4 a_S}{S(a_S)} \sqrt{(\rho_n \beta)^2+\frac{9}{64} \mu^2 \frac{S(a_S)^2}{D(a_S)^2} (a_S^2-a_n^2)^2}\\
tan(n \varphi_S) = -\frac{8}{3 \mu} \frac{\rho_n}{a_S^2-a_n^2} \frac{D(a_S)}{S(a_S)}
\label{eq:Fn_A_for_type_II}
\end{cases}
\end{align}
\subsubsection{Stability criterion.}
Having written the equilibrium condition, we shall now give a simple expression of the stability criterion. 
We shall use the general form \eqref{eq:equil_general} of the equilibrium condition.
To simplify the notations, Put\\
$\left. F'_a=\frac{dF}{da}\right|_{a=a_S}, \left. F'_\varphi=\frac{dF}{d\varphi}\right|_{a=a_S}, \left. G'_a=\frac{dG}{da}\right|_{a=a_S}, \text{ and } \left. G'_\varphi=\frac{dG}{d\varphi}\right|_{a=a_S}$.\\
To express the stability criterion, we use the classical method of the first-order expansion of the tangent system around the equilibrium point. We have, noting $a_S$ and $\varphi_S$ the values taken by $a$ and $\varphi$ at the equilibrium point:
\begin{align}
\begin{cases}
\dot{a} &=(a-a_S) F'_a+(\varphi-\varphi_S) F'_\varphi\\
\dot{\varphi}&=(a-a_S) G'_a+(\varphi-\varphi_S) G'_\varphi
\label{eq:equ2_19}
\end{cases}
\end{align}
The characteristic equation of this system is:
\begin{displaymath}
\begin{vmatrix}
F'_a-\lambda & F'_\varphi\\
G'_a & G'_\varphi-\lambda
\end{vmatrix}
=0
\end{displaymath}
that is,
$\lambda^2-(F'_a+G'_\varphi) \lambda+F'_a G'\varphi-F'_\varphi G'_a=0$.
The classical condition for this system to have a stable stationary solution is that the roots of the characteristic equation have negative real parts, i.e. that:
\begin{align}
\begin{cases}
\text{the sum of the real parts be negative, that is, } F'_a+G'_\varphi<0\\
\text{the product be positive, that is, } F'_a G'\varphi-F'_\varphi G'_a>0
\label{eq:equ2_18}
\end{cases}
\end{align}
In order to transform the first and second inequalities constituting the stability criterion in \eqref{eq:equ2_18}, we use the form \eqref{eq:equil_general} and, to simplify the notations, we put $\left. f=f(a_S), g=g(a_S), f'={\frac{df(a)}{da}} \right| _{a=a_S}, \left. g'={\frac{dg(a)}{da}} \right| _{a=a_S}$,\\
\paragraph{First inequality} ~\\
We have seen in \eqref{eq:equ2_18} that the first inequality linked to the stability criterion is:
\begin{equation}
F'_a+G'_\varphi <0,
\label{eq:equ2_28}
\end{equation}
which gives here: $A \sin(n\varphi_S) (f'-n j)+g'<0$. Replacing, in this expression, $A \sin(n\varphi_S)$ by its value deduced from \eqref{eq:equil_general_2}, we get: $-\frac{g}{f} (f'-n j)+g'<0$, which can be written:\\
\begin{equation}
f \left. \frac{d}{da}\left(\frac{g}{f}\right) \right|_{a=a_S, \varphi=\varphi_S}+n j \frac{g}{f}<0.
\label{eq:first_inequality_simplified}
\end{equation}
Given the definition of the averaged system \eqref{eq:equil_general}, this form will be much easier to manipulate than the original form \eqref{eq:equ2_28}. We can notice that the function h is not part of this inequality.
\paragraph{Second inequality} ~\\
We have seen in \eqref{eq:equ2_18} that the second inequality linked to the stability criterion is:
\begin{equation}
F'_a G'_\varphi-F'_\varphi G'_a>0,
\label{eq:equ2_29}
\end{equation}
Substituting in \eqref{eq:equ2_29} the developed values of $F'_a$, $F'_\varphi$, $G'_a$ and $G'_\varphi$ by their expressions as deduced from \eqref{eq:equil_general}, we get:
\begin{equation}
F'_a G'_\varphi-F'_\varphi G'_a=-(A f' \sin(n \varphi_S)+g') n A j sin(n\varphi_S)-n A f \cos(n\varphi_S) (A j' \cos(n\varphi_S)+h').
\label{eq:equ2_31}
\end{equation}
Being in a stationary condition, we can replace, in this expression, $A sin(n\varphi_S)$ and $A cos(n\varphi_S)$ by their expressions as deduced from \eqref{eq:equil_general_2}, thus obtaining:\\
\begin{equation}
F'_a G'_\varphi-F'_\varphi G'_a=n \left( -f' j \frac{g^2}{f^2}+g' j \frac{g}{f}-j' f \frac{h^2}{j^2}+h' f \frac{h}{j} \right)
\label{eq:equ2_32}
\end{equation}
Now, from \eqref{eq:A2_fn_of_aS_and_phiS}, expressing the derivative of $A^2$ with respect to $a_S$, we obtain:\\
\begin{equation}
\frac{1}{2} \frac{dA^2}{da_S}=\frac{g g' f-g^2 f'}{f^3}+\frac{h h' j-h^2 j'}{j^3}.\nonumber
\end{equation}
Therefore, by taking \eqref{eq:equ2_32} into account:
\begin{align}
\frac{1}{2} j f \frac{dA^2}{da_S} & = \frac{g g' j f^2-g^2 f' j f}{f^3}+\frac{h h' j^2 f-h^2 j j' f}{j^3} \nonumber \\
 & = \frac{g g' j}{f}-\frac{g^2 f' j}{f^2}+\frac{h h' f}{j}-\frac{h^2 j' f}{j^2} \nonumber\\
 & = \frac{1}{n} \left( F'_a G'_\varphi-F'_\varphi G'_a \right).\nonumber
\end{align}
And, because n is always positive, we can write the second inequality of the stability criterion as follows:
\begin{equation}
j f \frac{dA^2}{da}>0.
\label{eq:second_inequality_simplified}
\end{equation}
Because A is always positive by definition, we could use A instead of $A^2$, but the expression of A and its derivatives is much more intricate than that of $A^2$, and thus, we shall preferably use the form given in \eqref{eq:second_inequality_simplified}.
\paragraph{Application to Type B systems} ~\\
In \eqref{eq:first_inequality_simplified} and \eqref{eq:second_inequality_simplified}, substitute the expressions of $f$, $g$ and $j$ corresponding to the system of Type B, i.e.:\\
$f(a)=-\frac{S(a)}{4 \rho_n}$, $g(a)=-\beta a$ and $j(a)=-\frac{D(a)}{4 a \rho_n}$.\\
The first inequality \eqref{eq:first_inequality_simplified} becomes:
\begin{equation}
\frac{a}{S(a)} \frac{dS(a)}{da}-n \frac{D(a)}{S}-1<0
\label{eq:first_inequality_type_II}
\end{equation}
while the second inequality \eqref{eq:second_inequality_simplified} becomes: $S(a) D(a) \frac{dA^2(a)}{da}>0$. And because $A>0$, we can write:
\begin{equation}
S(a) D(a) \frac{dA(a)}{da}>0
\label{eq:second_inequality_type_II}
\end{equation}
Now take the case of the Type B system. Replacing the general expressions $S(a)$ and $D(a)$ by their known particular expressions given in \eqref{eq:S_D_and_S_D}, we can transform the first inequality \eqref{eq:first_inequality_type_II}, thus obtaining: $-1-\frac{1+\gamma a^2}{1+\gamma a^2}<0$, which is always true. In the same way, by transforming the second inequality \eqref{eq:second_inequality_type_II}, we get:
\begin{equation}
\left( \gamma a^2 - \left(\frac{2}{n^2-n \sqrt{n^2+4}+2}-1 \right) \right) \frac{dA}{da}>0,
\label{eq:second_stability_criterion_type_B}
\end{equation}
knowing that $\frac{dA}{da}$ and $\frac{dA^2}{da}$ are the same sign.\\
For $n>=3$, the expression $\gamma a^2 - \left(\frac{2}{n^2-n \sqrt{n^2+4}+2}-1 \right)$ can be approximated by $\gamma a^2-n^2-1$.\\
In conclusion, for the systems of Type B, the stability criterion is as of \eqref{eq:second_stability_criterion_type_B}, and can be approximated by
\begin{equation}
\left( \gamma a^2-n^2-1 \right) \frac{dA}{da}>0 .\\
\label{eq:second_stability_criterion_approx_type_B}
\end{equation}
\subsubsection{The $A(a_S)$ function for systems of Type B.}
\begin{figure}[h!]
\centering
\includegraphics[width=1.0\textwidth]{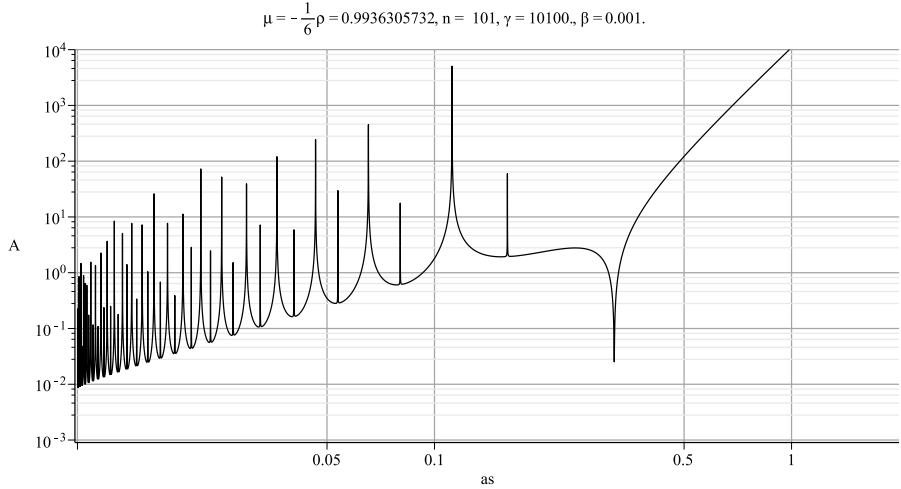}
\caption{$A(a_S)$ for a Type C system}
\label{fig:A(a_S) for a Type C system}
\end{figure}

\begin{figure}[h!]
\centering
\includegraphics[width=1.0\textwidth]{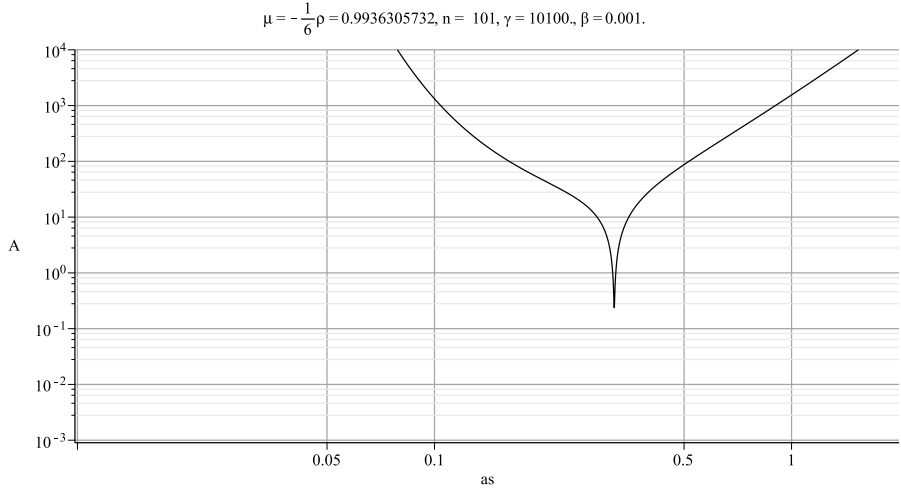}
\caption{$A(a_S)$ for a Type B system}
\label{fig:A(a_S) for a Type B system}
\end{figure}

\begin{figure}[h!]
\centering
\includegraphics[width=1.0\textwidth]{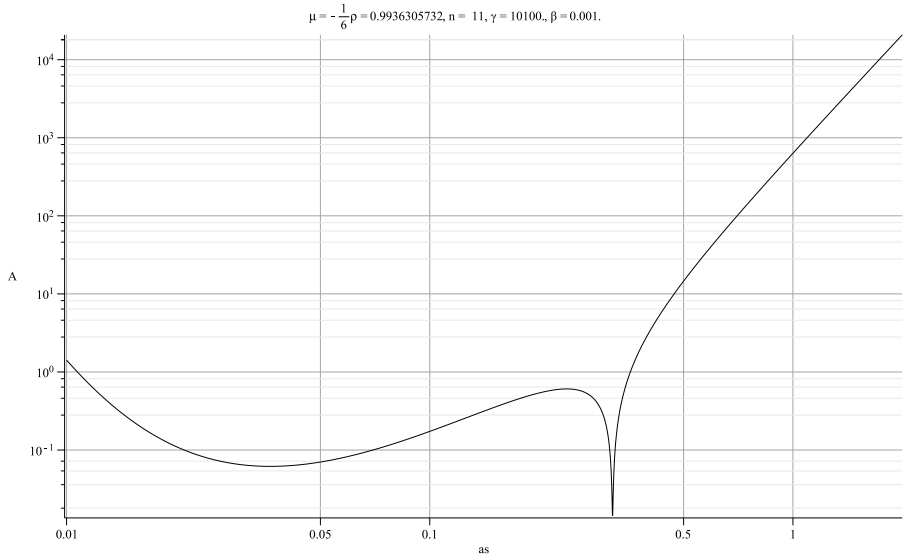}
\caption{$A(a_S)$ for a Type B system with more than one minimum for A}
\label{fig:A(a_S) for a Type B system with more than one minimum for A}
\end{figure}

When we studied the stability condition, we found an expression \eqref{eq:A2_fn_of_aS_and_phiS} giving $A$ as a function of $a_S$: the $A(a_S)$ function gives the amplitude of the external force as a function of the oscillator's amplitude at the equilibrium point $(a_S, \varphi_S)$, for a given n.\\
Then, when we studied the stability criterion, we found that, for systems of Type B, the stability criterion can be expressed very simply as a condition on the $A(a_S)$ function, as of equations \eqref{eq:second_stability_criterion_type_B} and \eqref{eq:second_stability_criterion_approx_type_B}.\\
Thus, the variation of the $A(a_S)$ function is worth being discussed.\\
First, we can see that when $a_S\rightarrow 0$ and when $a_S\rightarrow +\infty$, $\frac{a_S}{S(a_S)}\rightarrow +\infty$, and therefore $A(a_S)\rightarrow +\infty$. So the $A(a_S)$ function has at least one minimum on $R^+$.\\
Second, we can reasonably guess that we can have a local minimum around $a_S=a_n$.\\
To give an idea of the shape of the plot of $A(a_S)$, let's take an example: the case of the Type B system \eqref{eq:Fn_A_for_type_II}, where we have:
\begin{equation}
A(a_S) = \frac{4 a_S}{S(a_S)} \sqrt{(\rho_n \beta)^2+\frac{9}{64} \mu^2 \frac{S(a_S)^2}{D(a_S)^2} (a_S^2-a_n^2)^2}.
\label{eq:A_fn_of_aS}
\end{equation}
Being in the case of a system of Type B, we know the symbolic expressions of $S(a_S)$ and $D(a_S)$, given in \eqref{eq:S_D_and_S_D}. Substituting these expressions in \eqref{eq:A_fn_of_aS}, we can trace a plot of $A(a_S)$ for any numeric instance of the system parameters. We must keep in mind that our averaging calculus is only valid for small values of A, say $A<10$. However, we keep the plots as they are, for the sake of completeness and comparison between different models. Moreover, in some cases, the $A(a_S)$ function may exhibit more than one minimum, which is illustrated in figure \ref{fig:A(a_S) for a Type B system with more than one minimum for A}.\\
Take a typical case where $\gamma=10100$, $\mu=-\frac{1}{6}$ (pendulum case), $\nu=6.24*101$, $\omega_0=6.28$, $n_{crit}=100$, n=101 and $\beta=0.001$. Consequently, we have $\rho_n=\frac{6.24}{6.28}$.\\
In figure \ref{fig:A(a_S) for a Type C system}, we have a Type C system. In this model, we can see that the use of a $H$-function which is discontinuous introduces artifacts in the curve: the experiment and the numeric simulations using the averaged equations show that we don't have so many minima and stable regions in reality. However, the local minimum at $a_S\approx 0.4$ represents a real physical minimum and is usable for calculus and discussion.
~\\
In figure \ref{fig:A(a_S) for a Type B system}, we have a Type B system. In this model, we have a smooth curve for $A(a_S)$, because we used a smooth function for our $H$-function.\\
In figure \ref{fig:A(a_S) for a Type B system with more than one minimum for A}, we have another Type B system. In this model, we have a sharp minimum for the $A(a_S)$ function, due to the Duffing behaviour, plus a smooth minimum, due to the $S(a_S)$ function. We shall see later on how these two types of minimum lead to stable regimes.
\subsubsection{Bifurcations.}
\begin{figure}[h!]
\centering
\includegraphics[width=1.0\textwidth]{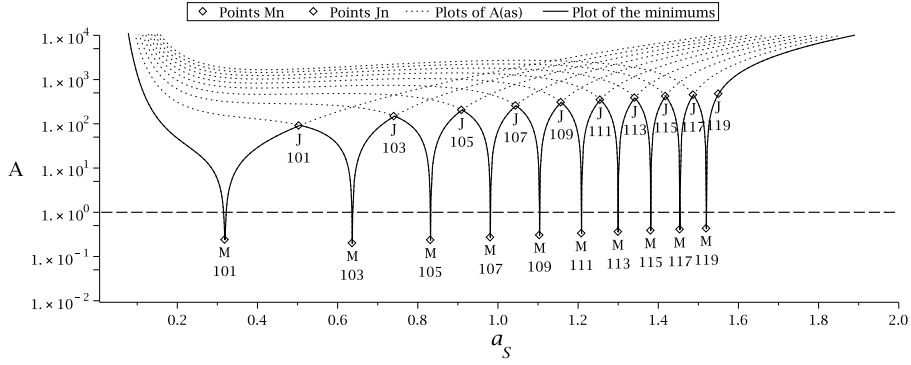}
\caption{$A(a_S)$ for a Type B system, with multiple values of n}
\label{fig:A_of_As_multiple_6_28_624}
\end{figure}

Until now, we discussed the averaged system on the basis of only one averaging calculus, based on one value of n. However, in the reality, we must take into account the fact that there is a plurality of values possible for n, each one corresponding to a value of the amplitude of the free oscillator, to an averaging calculus, and to an integral solution curve $U_n(a,\varphi)=0$. This way, we obtain a plurality of solutions, in the form of a plurality of integral curves $U_n(a,\varphi)=0$. These curves will overlap in the Van der Pol representations, and we shall have to decide which curve is the effective solution in a given domain of the Van der Pol plane.\\
For illustrative purpose, let's take the same typical case than for the representation of the  $A(a_S)$ curve of Type B in figure \ref{fig:A(a_S) for a Type B system}, where $\gamma=10100$, $\mu=-\frac{1}{6}$, $\omega_0=6.28$, nu=624, $\beta=0.001$.\\
As we took $\mu$ negative, we have to take $n>n_{crit}+1=100$, i.e. $n\geqslant 101$. Let's plot the $A(a_S)$ curves for n=101, 103, ..., 119. We get a series of overlapped curves (see figure \ref{fig:A_of_As_multiple_6_28_624}), and we are interested in the local minimums indicated by letters $M_{101}...M_{111}$. For each minimum $M_i$, we shall spot a point $N_i$ having an abscissa slightly greater than that of $M_i$. As we saw before, these points represent stable solutions to the averaged system of equations, because they are on an ascending part of the curve $A(a_S)$.\\
Because the averaged method gives better results when the perturbation is small, we shall (in a first approach) keep the parts of the overlapped curves which are the lowest in ordinates, i.e. corresponding to the lowest values of A, the intensity of the perturbation.\\
In figure \ref{fig:A_of_As_multiple_6_28_624} are also represented the points $J_n$, which are the intersections of curves $A(a_S)$ for two consecutive values of n: n and n+2.
Denote by $A_n$ the ordinate of $M_n$. From equation \eqref{eq:A_fn_of_aS}, we have:
\begin{equation}
A_n=A(a_n)=\frac{4 a_n}{S(a_n)} \rho_n \beta.
\label{eq:An}
\end{equation}
And as the ordinate of $M_n$ increases as the system parameter A increases, we can see that the number of stable solutions to the original non-averaged system \eqref{eq:eq2_13} and \eqref{eq:eq2_14}, i.e. the number of $A(a_S)$ curves cut by a given line $A=Cte$, increases as A increases, constituting the phenomenon of bifurcation.\\
To have a more precise indication of the limits (in abscissa $a$) of the region where the averaging method is valid for a given value of $n$, we must find a symbolic expression for the coordinates of the points $J_n$. Denote by $\overset{q}{A}(a_S)$ the curve representing the function $A(a_S)$ when $n$ has the value $q$. Then $J_n$ is the intersection of $\overset{n}{A}(a_S)$ and $\overset{n+2}{A}(a_S)$.\\
The abscissa $x$ of $J_n$ satisfies the equation \eqref{eq:A_fn_of_aS} with $n$ for $\overset{n}{A}(a_S)$ as well as this same equation with $n+2$ for $\overset{n+2}{A}(a_S)$. We get, after a few transformations:
\begin{equation}
\lambda^2 \beta^2 \left(\frac{1}{n^2}-\frac{1}{(n+2)^2}-2 \right)=\frac{9}{64} \: \mu^2 \left( \frac{S(x)}{D(x)} \right)^2 (a_n^2-a_{n+2}^2) (2 x^2-a_n^2-a_{n+2}^2),
\label{eq:x_an_anp2}
\end{equation}
with $\lambda=\frac{\nu}{\omega_0}$.\\
Replacing $a_n$ and $a_{n+2}$ by their developed expressions as given in equation \eqref{eq:definition_of_a_n}, we get:
\begin{equation}
x^2=\frac{2}{3 \mu} \left(\frac{\lambda^2}{n^2}+\frac{\lambda^2}{(n+2)^2}-2 \right)+\frac{8 \beta^2}{3 \mu} \left( \frac{D(x)}{S(x)} \right).
\end{equation}
Assuming that, for x in the interval $[a_n, a_n+2]$, $S(x) \approx S\left(\xi \right)$ and $D(x) \approx D\left(\xi \right)$, with $\xi=\frac{a_n+a_{n+2}}{2}$, and remarking that $a_n^2+a_{n+2}^2=\frac{4}{3 \mu} \left(\frac{\lambda^2}{n^2}+\frac{\lambda^2}{(n+2)^2}-2 \right)$, we finally get:
\begin{equation}
a_{n, n+2}= \sqrt{a_n^2+a_{n+2}^2+\frac{8}{3 \mu} \lambda^2 \beta^2 \left( \frac{D(\xi)}{S(\xi)} \right)^2},
\label{eq:an_anp2}
\end{equation}
where we denote by $a_{n, n+2}$ the abscissa of $J_n$, intersection of the curves $\overset{n}{A}(a_S)$ and $\overset{n+2}{A}(a_S)$.
\subsection{The rectangular Van der Pol representation.}
\begin{figure}[h!]
\centering
\includegraphics[width=1.0\textwidth]{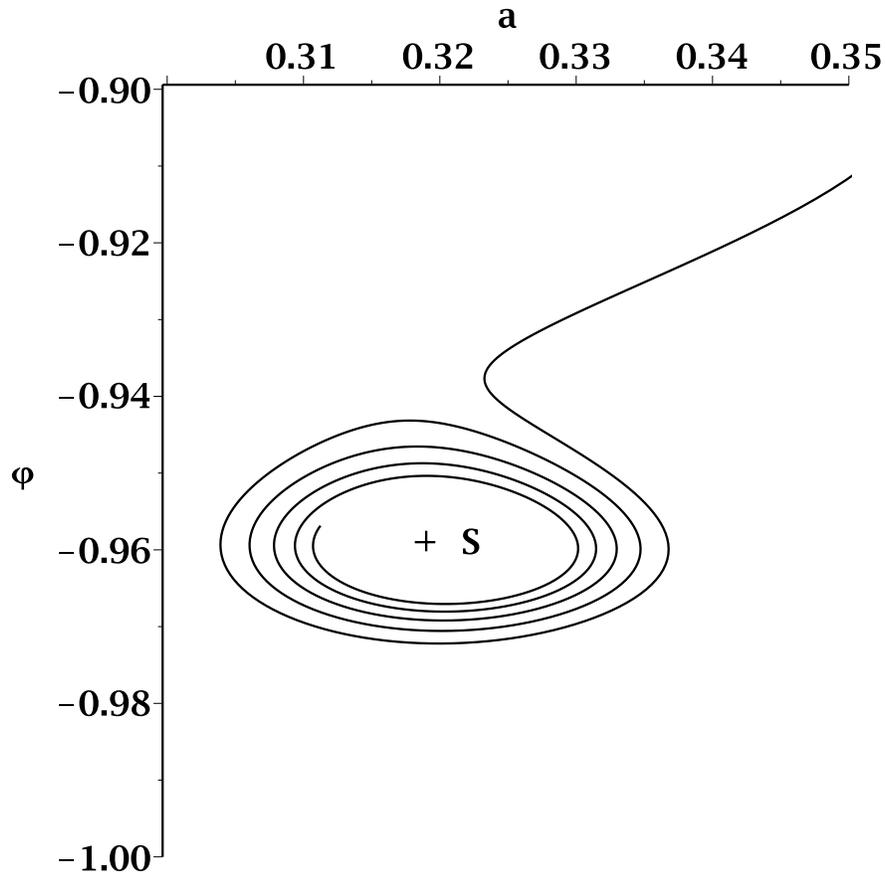}
\caption{Integral curve with stable equilibrium at point S. Parameters: $\nu=624$, $\omega_0=6.28$, n=101, $\beta=0.001$, $\mu=-\frac{1}{6}$, A=0.534, $\gamma=10100$. Initial conditions: $a_0=0.45$, $\varphi_0=0.$}.
\label{fig:Courbe 2}
\end{figure}

As we previously mentioned, we are naturally induced to use a Van der Pol representation, with $a$ as absciss$a$ and $\varphi$ as ordinates. In figure \ref{fig:Courbe 2}, we have an example of an integral curve in this representation. The parameters are the same as previously. The integral curve winds into a spiral leading up to a stable equilibrium represented as point S. This rectangular representation is useful to assess various probabilities in terms of areas in the plane. As the probability is uniformly distributed against the abscissa and the ordinates, we can carry out the calculus of areas without any risk of giving an excess weight to a given region.
\subsection{The polar Van der Pol representation.}
\begin{figure}[h!]
\centering
\includegraphics[width=1.0\textwidth]{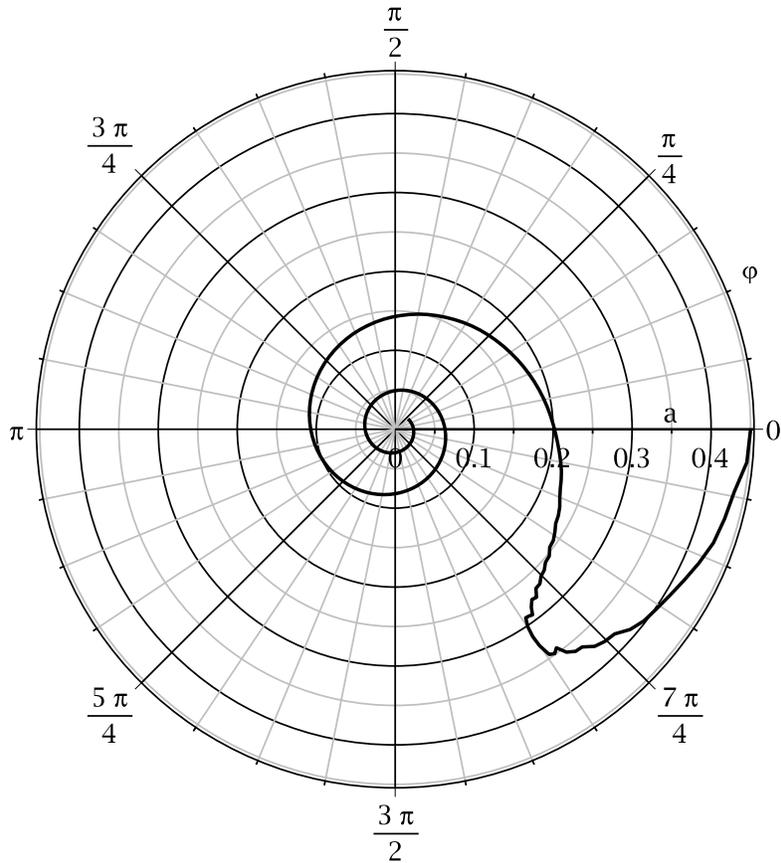}
\caption{Polar representation: integral curve winding up around the origin. Radius is a, argument is $\varphi$. Parameters: $\nu=624$, $\omega_0=6.28$, n=101, $\beta=0.001$, $\mu=-\frac{1}{6}$, A=0.534, $\gamma=10100$. Initial conditions: $a_0=0.45$, $\varphi_0=0.$}
\label{fig:Winding spiral}
\end{figure}

\begin{figure}[h!]
\centering
\includegraphics[width=1.0\textwidth]{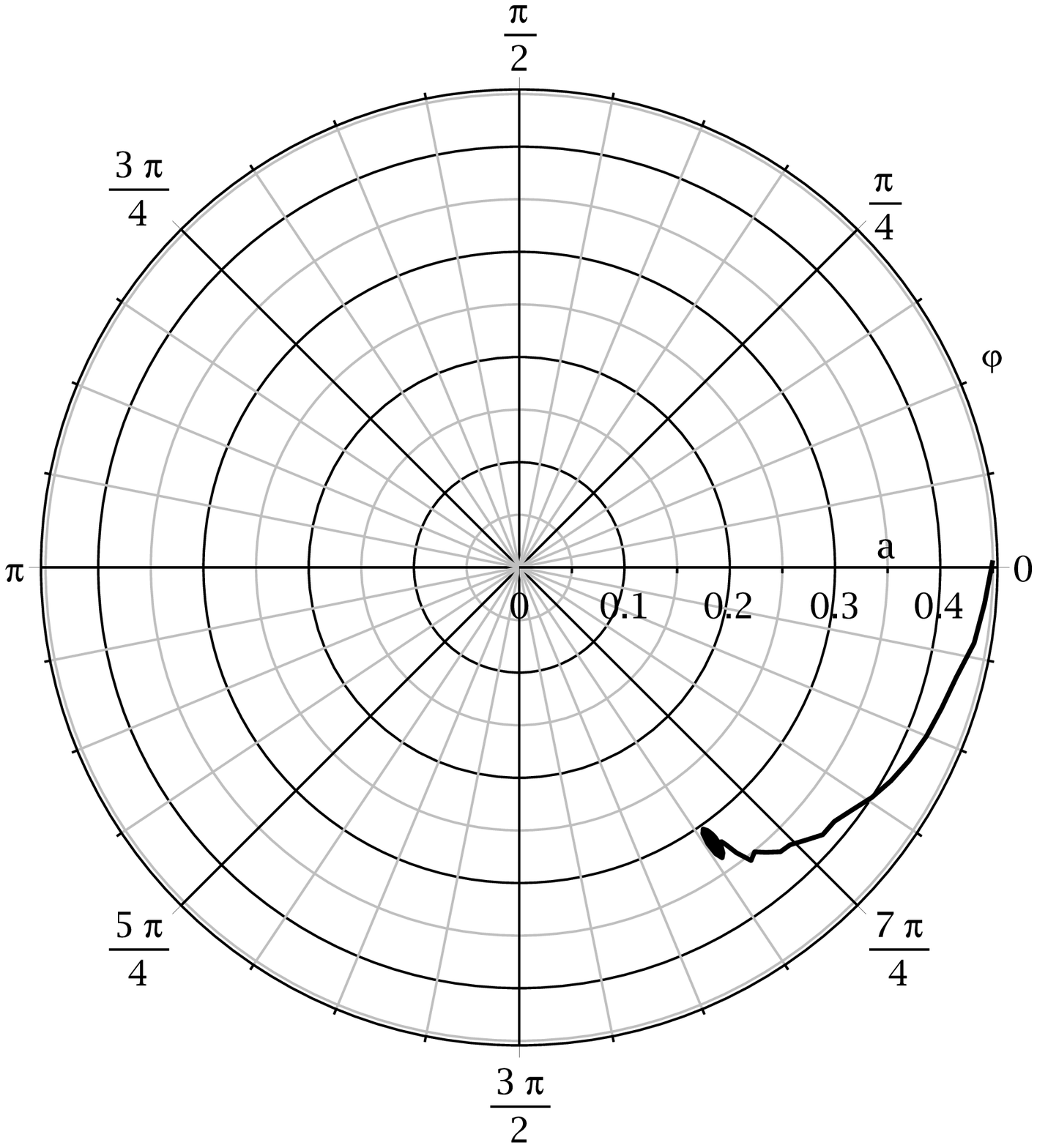}
\caption{Polar representation: integral curve winding up around a stable solution, not at the origin.}
\label{fig:Winding spiral_not_center}
\end{figure}

\begin{figure}[h!]
\centering
\includegraphics[width=1.0\textwidth]{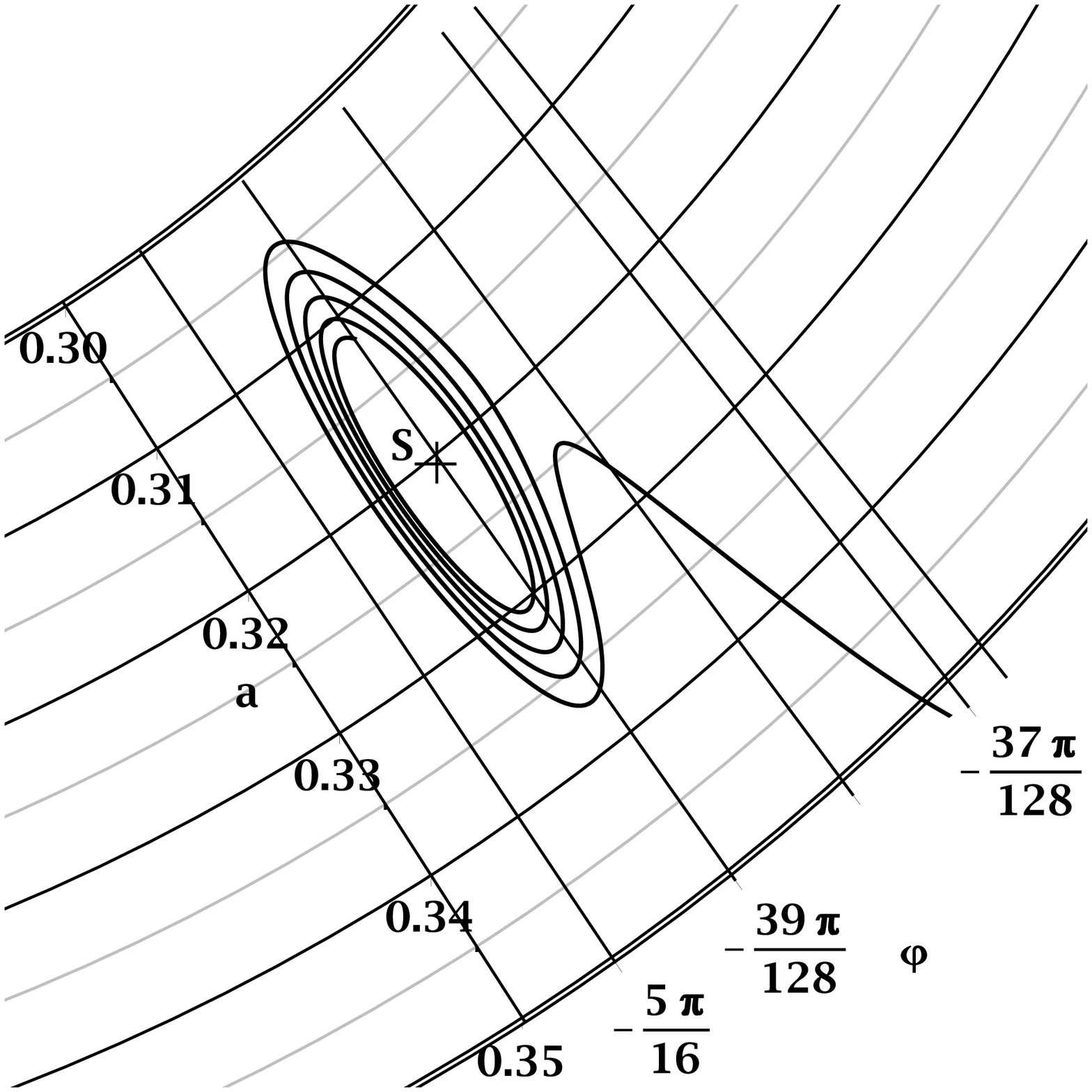}
\caption{Polar representation: integral curve winding up around a stable solution represented as point S, not at the origin. Zoomed view.}
\label{fig:Winding spiral_not_center_zoomed}
\end{figure}

We also use a polar Van der Pol representation, i.e. $a$ as radius and $\varphi$ as angle. This polar representation is useful to represent plots having a periodicity with respect to $\varphi$ and plots in which $\varphi$ varies globally by more than a given finite interval. In particular, we can notice that our averaged system of equations \eqref{eq:averaged_system} is invariant by the transformation $\varphi \rightarrow \varphi+\frac{\pi}{2 n}$. Therefore, in the polar Van der Pol representation, the plot of the integral curves will be invariant by a rotation of angle $\frac{\pi}{2 n}$, and we will be able to obtain the entirety of the plot by duplication and rotation of only one set of solution curves, located in a given sector. For instance, with parameters identical to those we used about the bifurcations, figure \ref{fig:Winding spiral} is a polar Van der Pol representation of the integral curve computed using a Runge-Kutta Fehlberg method that produces a fifth-order accurate solution, with initial conditions $a_0=0.45$ and $\varphi_0=0.018$. The solution winds up around the origin, and could not be entirely represented in a rectangular plot, because although $a(t)$ remains finite, $\varphi(t)$ can become big when $t$ increases.\\
With the same parameters, if we take the initial conditions $a_0=0.45$, $\varphi_0=0.01473684211$, we get a completely different behaviour, as represented in figure \ref{fig:Winding spiral_not_center}: the movement enters a spiral winding up around a stable solution not located at the origin.
The zoomed view in figure \ref{fig:Winding spiral_not_center_zoomed} shows the detail in polar coordinates. This represents the same data as in figure \ref{fig:Courbe 2}, which was in rectangular coordinates.
\section{Third step: approximate analytical solution for the damped Duffing oscillator.}
In this section, we shall present an approximate closed-form solution of the averaged system of equations \eqref{eq:averaged_system}. We shall use an integrating factor, which will lead us to an implicit equation of the integral curve (independent of the reduced time $\tau$). We shall then discuss the validity of the approximation.\\
\subsection{Integrating factor.}
Let's recall our averaged system \eqref{eq:averaged_system}:
\begin{align}
 \begin{cases}
  \dot{a} &= -\frac{A}{4 \rho} S(a) \sin(n \varphi)-\beta a \\
  \dot{\varphi}&= \frac{3\mu}{8 \rho}(a^2-a_n^2)-\frac{A}{4 a \rho_n} \cos(n \varphi) D(a)
 \end{cases}
\label{eq:29}
\end{align}
We shall write \eqref{eq:29} under a more compact form, by defining two functions $F$ and $G$ of $a$ and $\varphi$ as follows:
\begin{align}
 \begin{cases}
  F(a,\varphi) &= -\frac{A}{4 \rho} S(a) \sin(n \varphi)-\beta a \\
  G(a,\varphi)&= \frac{3\mu}{8 \rho}(a^2-a_n^2)-\frac{A}{4 a \rho_n} \cos(n \varphi) D(a)
 \end{cases}
\label{eq:def_F_and_G}
\end{align}
Hence:
\begin{align}
 \begin{cases}
  \dot{a}&=F(a, \varphi) \\
  \dot{\varphi}&=G(a, \varphi)
 \end{cases}
\end{align}
that is,
\begin{equation}
  G(a, \varphi) da + F(a, \varphi) d\varphi=0.
\label{eq:I_diff}
\end{equation}
This is generally not an exact differential, but if we multiply the whole equation by an integrating factor, i.e an appropriate function $I(a, \varphi)$ such that there exists a function $U(a, \varphi)$ such that $I(a, \varphi) G(a, \varphi) da +I(a, \varphi) F(a, \varphi) d\varphi=dU(a, \varphi)$, equation \eqref{eq:I_diff} becomes $dU(a, \varphi)=0$, and the solution is $U(a, \varphi)=constant$. The new expression $dU(a, \varphi)$ is an exact differential. A necessary condition for this to be possible is that $\frac{\partial}{\partial \varphi}(I(a, \varphi) G(a, \varphi))=-\frac{\partial}{\partial a}(I(a, \varphi) F(a, \varphi))$. Replacing $F$ and $G$ by their definition expressions from \eqref{eq:def_F_and_G}, and developing, we get:
\begin{align}
&\left( \frac{A n}{4 a \rho_n} D(a) \sin(n \varphi)-\frac{A}{4 \rho_n} \frac{dS(a)}{da} sin(n \varphi)-\beta \right) I(a, \varphi)= \nonumber \\
&\left( \frac{A}{4 \rho_n} S(a) \sin(n \varphi)+\beta a  \right) \frac{\partial}{\partial a}I(a, \varphi)-\left( \frac{3}{8}\frac{\mu}{\rho_n}(a^2-a_n^2)-\frac{A}{4 a \rho_n} D(a) \cos(n \varphi) \right) \frac{\partial}{\partial \varphi}I(a, \varphi)
\label{eq:integr_factor_2_var}
\end{align}
We shall not try to find an integrating factor as a function of $a$ and $\varphi$; instead, we shall search $I$ as a function of $a$ only. By so doing, the integrating factor becomes approximatively findable in closed-form. Let $I(a)$ be the unknown integrating factor. 
For the sake of clarity, we shall represent the following functions of a: $S(a), D(a), I(a), \frac{dS(a)}{da}, \frac{dD(a)}{da}, \frac{dI(a)}{da}$ respectively by $S, D, I, S', D', I'$. 
Equation \eqref{eq:integr_factor_2_var} becomes:
\begin{equation}
\frac{A}{4 \rho_n} \sin(n \varphi) \left( \frac{n D}{a} I-S' I-S I' \right)=\beta \left(a I' +I \right)
\end{equation}
and, due to the fact that $a$ and $\varphi$ are independent variables, this is possible only if:
\begin{align}
 \begin{cases}
  I \frac{A n}{4 a \rho_n} D-I' \frac{A}{4 \rho_n} S-I \frac{A}{4 \rho_n} S'\equiv 0 \\
  I' \beta a+I \beta \equiv 0
 \end{cases}
\end{align}
that is, we obtain the two following equations:
\begin{eqnarray}
  \frac{n D}{a}I-I S'-I' S \equiv 0 \\
\label{eq:I_I_prime_1}
  a I'+I \equiv 0.
\label{eq:I_I_prime_2}
\end{eqnarray}
Equation \eqref{eq:I_I_prime_2} gives $I(a)=\frac{k}{a}$, where $k$ is a real constant. And as the result is equivalent for any non-null value of $k$, as is expressed by $k U(a, \varphi)=0$, we shall take $k=1$ to simplify the writing. By substituting this expression of $I(a)$ into equation \eqref{eq:I_I_prime_1}, we get: $\frac{n}{a^2} D+\frac{n}{a^2} S-\frac{1}{a} S'\equiv0$, that is,
\begin{equation}
n D+S-a S'\equiv0.
\label{eq:D_S_S_prime}
\end{equation}
This equation is not strictly verified by the functions $S(a)$ and $D(a)$, but we can remark that generally, those two functions do not vary very much in the region of the spiral leading to the stable stationary solution. We shall then be able to use one of the following three methods to replace $S$ or $D$ by another function so as to satisfy \eqref{eq:D_S_S_prime}:
\begin{itemize}
\item Keeping the original definition of $S(a)$, and replacing $D(a)$ by $\frac{a S'-S}{n}$;
\item Keeping the original definition of $D(a)$, and replacing $S(a)$ by $n a \int \frac{D}{a^2}da+C_1 a$, where $C_1$ is a constant;
\item Replacing $S$ by an affine function of $a$ and $D$ by a constant function. Let $a_S$ be the value of $a$ in the stationary solution, represented by the point S in figures \ref{fig:Courbe 2} and \ref{fig:Winding spiral_not_center_zoomed}. Let's take $S'(a_S)=\frac{n D(a_S)+S(a_S)}{a_S}$. We then have $S(a)=S(a_S)+S'(a_S) (a-a_S)$ and $S(a)-a S'(a)=S(a_S)-a_S S'(a_S)=Cte=n D(a)$.
\end{itemize}
The numerical simulations show that the two first methods are approximately equivalent and lead to satisfactory results for oscillators of Type B. Supposing that we now have an $S(a)$ and a $D(a)$ functions satisfying equation \eqref{eq:D_S_S_prime}, we can now go on and compute the function $U(a, \varphi)$ using one of the two following formulas:
\begin{eqnarray}
\frac{\partial U}{\partial a}=I(a) G(a, \varphi) \\
\label{U_I_G}
\frac{\partial U}{\partial \varphi}=-I(a) F(a, \varphi).
\label{U_I_F}
\end{eqnarray}
Let's pick \eqref{U_I_F}. Replacing $F(a, \varphi)$ by its definition expression from \eqref{eq:def_F_and_G}, we get:
\begin{equation}
U(a, \varphi)=-\frac{A}{4 \rho_n} \frac{\cos(n \varphi)}{n} \frac{S}{a}+\beta \varphi+L(a),
\label{eq:U_a_phi}
\end{equation}
where $L(a)$ is a function of a to be determined using equation \eqref{U_I_G}.\\
Calculating $\frac{\partial U}{\partial a}$ from equation \eqref{eq:U_a_phi} and substituting the result into equation \eqref{U_I_G}, we get:
\begin{equation}
\frac{A}{4 \rho_n} \left( \frac{1}{n} (S-S' a)+D \right) \cos(n \varphi)=\frac{3}{8} \frac{\mu}{\rho_n} a (a^2-a_n^2)-a^2 L'(a).
\label{eq:S_D_L_prime}
\end{equation}
As $a$ and $\varphi$ are independent variables, this is possible only if:
\begin{align}
 \begin{cases}
  S-S' a+n D \equiv 0 \\
  \frac{3}{8} \frac{\mu}{\rho_n} (a^2-a_n^2)-a L'(a) \equiv 0.
 \end{cases}
\end{align}
We have already encountered and discussed the first condition previously. The second condition gives: $L'(a)=\frac{3}{8} \frac{\mu}{\rho_n} \left( a-\frac{a_n^2}{a} \right)$. Hence $L(a)=\frac{3}{8} \frac{\mu}{\rho_n} \left( \frac{a^2}{2}-a_n^2 \ln(a) \right)+C_2$, where $C_2$ is a constant.\\
Finally, substituting this expression of $L(a)$ into \eqref{eq:U_a_phi}, we get:
\begin{equation}
U(a, \varphi)=-\frac{A S(a)}{4 n \rho_n} \frac{\cos(n \varphi)}{a}+\beta \varphi+\frac{3}{8} \frac{\mu}{\rho_n} \left( \frac{a^2}{2}-a_n^2 \ln(a) \right)+C_3
\label{eq:complete_U}
\end{equation}
where $C_3$ is a constant and with $n D(a) \equiv a S'(a)-S(a)$.
\subsection{Implicit equation of the integral curves.}
\begin{figure}[h!]
\centering
\includegraphics[width=1.0\textwidth]{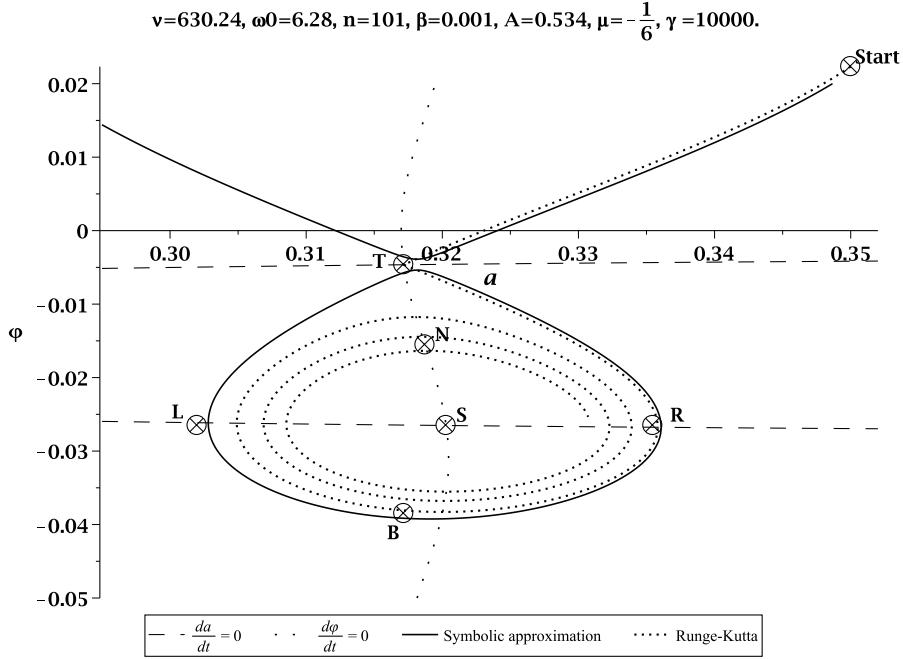}
\caption{Symbolic approximation vs numeric (Runge-Kutta) solution to the averaged system.}
\label{fig:Symbolic ovoid and Runge-Kutta solution}
\end{figure}

From equation \eqref{eq:complete_U}, we deduce the implicit equation of the integral curve beginning at the initial condition $(a=a_0, \varphi=\varphi_0)$: $U(a, \varphi)=U(a_0, \varphi_0)$. Developing and eliminating the constant $C_3$, we get:
\begin{align}
&\beta (\varphi-\varphi_0)-\frac{A}{4 n \rho_n} \left( \frac{S(a) \cos(n \varphi)}{a}-\frac{S(a_0) \cos(n \varphi_0)}{a_0} \right)+ \nonumber \\
& \qquad \frac{3}{8} \frac{\mu}{\rho_n} \left( \frac{a^2-a_0^2}{2}-a_n^2 (\ln(a)-\ln(a_0)) \right)=0.
\label{eq:integral_curve_implicit_equation}
\end{align}
In figure \ref{fig:Symbolic ovoid and Runge-Kutta solution}, the implicit equation \eqref{eq:integral_curve_implicit_equation} is represented by a solid line, with a typical set of system parameters. The corresponding numeric solution is represented as a dotted line. Dashed lines represent the locus of the condition $\frac{da}{dt}=0$, and a dotted line represents the locus of the condition $\frac{d\varphi}{dt}=0$.
Depending on the initial conditions, the symbolic approximation is composed of either a single curve surrounding the ovoid region, or two curves, one of which surrounds the ovoid region, the other following the numeric solution before arriving at the ovoid, and then going away from the ovoid. In figure \ref{fig:Symbolic ovoid and Runge-Kutta solution} are also represented a number of points which will be discussed in the next sections.
\subsection{Group of ovoids inside an annulus for a given value of n.}
In figure \ref{fig:Symbolic ovoid and Runge-Kutta solution}, we have represented one ovoid in rectangular Van der Pol representation, using the symbolic implicit solution. But as our averaged system of equations is invariant by a rotation of angle $\frac{2 \pi}{n}$, we shall represent this ovoid in a polar system of coordinates, with $a$ as radius and $\varphi$ as argument. Because of said invariance, we shall duplicate $n$ times the ovoid along a circle centred at the centre of coordinates, by successive rotations of the initial ovoid by an angle of $\frac{2 \pi}{n}$.
\begin{figure}[h!]
\centering
\includegraphics[width=1.0\textwidth]{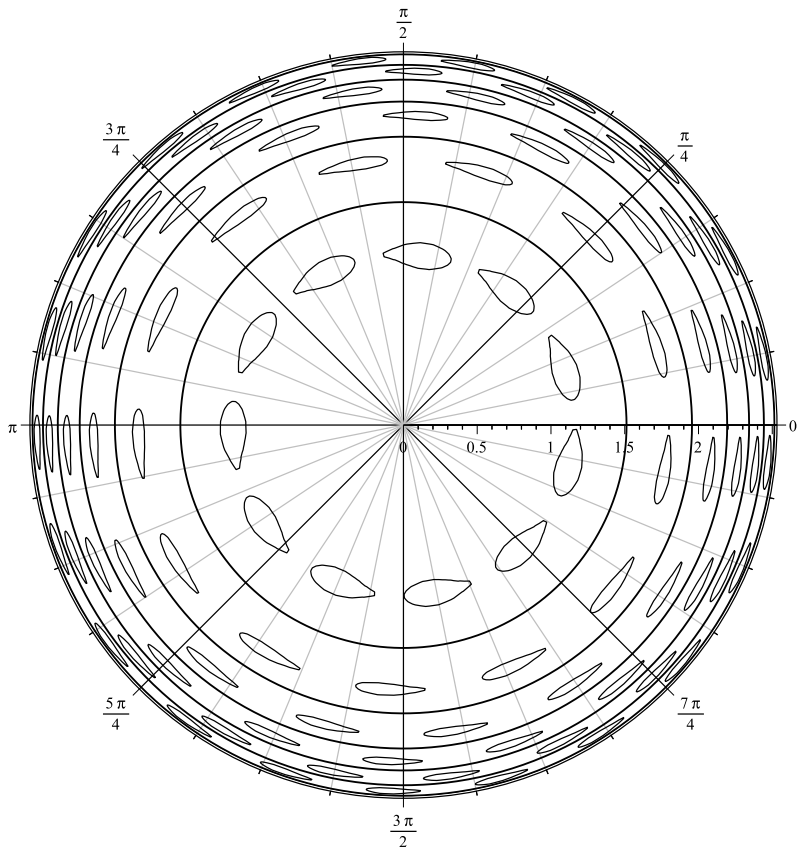}
\caption{Multiple ovoids in 6 consecutive annuli.}
\label{fig:multiple_ovoids_in_multiple_annuli}
\end{figure}
In figure \ref{fig:multiple_ovoids_in_multiple_annuli}, the contents of 6 adjacent annuli is represented, as well as the circles of radii $a_{n, n+2}$ obtained from equation \eqref{eq:an_anp2}. Those circles delimit, for each value of n, an upper bound of the validity region of the averaging method that we used.
\section{Fourth step: capture probability.}
In this section, we shall use the closed-form implicit equation \eqref{eq:integral_curve_implicit_equation} to assess the area of the ovoid basin, and compare it to the area of the annulus containing the ovoid. This will lead us to a symbolic formula giving the capture probability for initial conditions inside said annulus.\\
Let's call ``angular diameter'' the maximum distance between two points of the ovoid, said points having the same radius in the polar Van der Pol representation. And we shall call ``radial diameter'' the maximum distance between two points of the ovoid, said points having the same angle. We shall calculate those two diameters and deduce a value of the area of the ovoid.
\subsection{Angular positions of points S and T in figure \ref{fig:Symbolic ovoid and Runge-Kutta solution}.}
In figure \ref{fig:Symbolic ovoid and Runge-Kutta solution}, we represent four points around the ovoid region, called T (at coordinates $(a_T, \varphi_T)$) at the top, R (at coordinates $(a_R, \varphi_R)$) at the right, B (at coordinates $(a_B, \varphi_B)$) at the bottom, and L (at coordinates $(a_L, \varphi_L)$) at the left. T and B correspond to $\frac{d\varphi}{dt}=0$, while L and R correspond to $\frac{da}{dt}=0$. As of figure \ref{fig:Symbolic ovoid and Runge-Kutta solution}, those points are not exactly on the perimeter of the ovoid, because their location is calculated in an approximate way, as described hereafter. We also represent point S (at coordinates $(a_S, \varphi_S)$), which is the stable-equilibrium solution to the averaged system \eqref{eq:averaged_system}, while point T represents the unstable-equilibrium solution. As T and S both satisfy the equilibrium condition $(\dot{a}=0, \dot{\varphi}=0$), we can calculate the position of these points by getting back to the averaged system \eqref{eq:averaged_system} with $a=a_S$, $\varphi=\varphi_S$, $\dot{a}=0$ and $\dot{\varphi}=0$. We have:
\begin{align}
 \begin{cases}
  0 &= -\frac{A}{4 \: \rho_n} \ S(a_S) \sin(n \varphi)-\beta \: a_S \\
  0 &= \frac{3 \: \mu}{8 \: \rho_n}(a_S^2-a_n^2)-\frac{A}{4 \: a_S \: \rho_n} \cos(n \varphi) \: D(a_S)
 \end{cases}
\end{align}
Hence, with $a_S\approx a_n$, $S(a_S) \approx S(a_n)$ and $D(a_S) \approx D(a_n)$:
\begin{align}
\begin{cases}
a_S & \approx \frac{A S(a_n)}{4 \rho_n \beta}\\
sin(n \varphi_S) & = -\frac{4 \rho_n \beta a_n}{A S(a_n)}
\label{eq:giving_a_S_and_phi_S}
\end{cases}
\end{align}
Putting $S_n=S(a_n)$, we get, in the interval $\left[-\frac{\pi}{n},\frac{\pi}{n} \right]$, two possible values for $\varphi_S$: $-\frac{1}{n} \arcsin\left(\frac{4 \beta \rho_n a_n}{A S_n}\right)$ and $-\frac{\pi}{n}+\frac{1}{n} \arcsin\left(\frac{4 \beta \rho_n a_n}{A S_n}\right)$. Because S and T both represent equilibrium conditions, both $\varphi_S$ and $\varphi_T$ satisfy equations \eqref{eq:giving_a_S_and_phi_S}. Consider the case of a Type B system, where $a_T<a_n<a_S$, which is true, as we discussed earlier, if point $(a_T,A(a_T))$ is on a decreasing part of the plot representing A as a function of $a$ (hence in an unstable-equilibrium region), while point $(a_S,A(a_S))$ is on an increasing part of said plot (hence a stable-equilibrium region). We then have:
\begin{align}
\begin{cases}
\varphi_S & = -\frac{\pi}{n}+\frac{1}{n} \arcsin\left(\frac{4 \beta \rho_n a_n}{A S_n}\right) \\
\varphi_T & = -\frac{1}{n} \arcsin\left(\frac{4 \beta \rho_n a_n}{A S_n}\right).
\label{eq:expression_of_phi_S_and_phi_T}
\end{cases}
\end{align}
\subsection{Angular position of point B. Angular diameter of the ovoid.}
Consider the points T and B in figure \ref{fig:Symbolic ovoid and Runge-Kutta solution}. Those two points delimit the angular extension of the ovoid. As we search an expression for the angular diameter of the ovoid, i.e. $\varphi_T-\varphi_B$, and know the expression of $\varphi_T$, we have to find an expression for $\varphi_B$. \\
We shall consider that, in the rectangular Van der Pol representation, the ovoid has a shape approximatively symmetric with respect to the vertical. Hence we shall write:
\begin{equation}
a_B \approx a_T,
\label{eq:aB_aS}
\end{equation}
and consequently, $S(a_B) \approx S(a_T)$.
Point B is on the integral curve passing through point T. The equation of this integral curve is: $U(a, \varphi)=U(a_T, \varphi_T)$, i.e. equation \eqref{eq:integral_curve_implicit_equation}, where we replace $(a_0, \varphi_0)$ by $(a_T, \varphi_T)$. Let's write that point B is on this curve. Substituting the coordinates of point B for the current coordinates $(a, \varphi)$ in this equation, and taking \eqref{eq:aB_aS} into account:
\begin{equation}
\beta (\varphi_B-\varphi_T)-\frac{A S(a_T)}{4 n \rho_n a_T} \left( \cos(n \varphi_B)-\cos(n \varphi_T) \right) \approx 0.
\label{eq:Implicit_curve_through_phi_T}
\end{equation}
From equations \eqref{eq:expression_of_phi_S_and_phi_T}, we know the expression of $\varphi_T$. Furthermore, as point T is an equilibrium state, $(a_T,\varphi_T)$ satisfies the system \eqref{eq:giving_a_S_and_phi_S} by substituting point T for point S. Hence we have :
\begin{align}
\begin{cases}
a_T & \approx \frac{A S(a_n)}{4 \rho_n \beta}\\
sin(n \varphi_T) & = -\frac{4 \rho_n \beta a_n}{A S(a_n)}.
\label{eq:giving_a_T_and_phi_T}
\end{cases}
\end{align}
Considering that function S(a) varies slowly inside the ovoid and that $a_T$ is close to $a_n$, we put $S(a_T) \approx S(a_n)$; thus, by substituting $S(a_n)$ for $S(a_T)$ and $-\frac{\beta }{n sin(n \varphi_T)}$ for $ \frac{A S(a_n)}{4 \rho_n \beta a_n}$ in equation \eqref{eq:Implicit_curve_through_phi_T}, we get:
\begin{equation}
n \varphi_B-n \varphi_T+\frac{1}{sin(n \varphi_T)} \left( \cos(n \varphi_B)-\cos(n \varphi_T) \right) \approx 0.
\label{eq:giving_phi_T}
\end{equation}
$\varphi_T$ being known, this is a transcendental equation in $\varphi_B$, which we shall solve approximately as follows. \\
Put $x=n \varphi_B$ and $x_0=n \varphi_T$. Equation \eqref{eq:giving_phi_T} becomes :
\begin{equation}
\sin(x_0) \; (x-x_0)=\cos(x)-\cos(x_0),
\label{eq:x_x0}
\end{equation}
which means that we are searching the intersection of the curve $z(x)=\cos(x)$ with the line $z(x)=\sin(x_0) \; (x-x_0)+cos(x_0)$. A first solution is $x=x_0$, i.e. $\varphi_B=\varphi_T$. What we search is the other solution. $x_0$ is in the limited range $[0, \pi /2]$. Over this range, we approximate the curve implicitly given by \eqref{eq:x_x0} by using the curve giving x explicitly as a function of $x_0$ as follows:
\begin{equation}
x(x_0)=-\frac{\pi}{2}-3 \arcsin\left(1+\frac{2}{\pi} x_0\right),
\label{eq:x_x0_explicit}
\end{equation}
Getting back to the initial problem, where we had put $x=n \varphi_B$ and $x_0=n \varphi_T$, we get:
\begin{equation}
\varphi_B \approx -\frac{1}{n} \left(\frac{\pi}{2} +3 \arcsin\left(1+\frac{2 n}{\pi} \varphi_T\right)\right).
\label{eq:x_phiB}
\end{equation}
We remark that if $n \varphi_T=-\frac{\pi}{2}$, then $n \varphi_B=-\frac{\pi}{2}$. In this case, both solutions $\varphi_B$ and $\varphi_T$ are identical, and the ovoid is reduced to a single point.
\subsection{Capture probability with initial condition $a=a_n$.}
To assess this probability, we shall calculate the angular diameter of the ovoid, in the rectangular Van der Pol representation, that is the distance TB, and form the ratio of this distance to the distance between the centres of two contiguous ovoids.\\
Now that we know the expression giving $\varphi_B$, we can calculate the angular diameter $D_{ang}=TB=a_n (\varphi_T - \varphi_B)$ of the ovoid, along the radius $a_n$:
\begin{equation}
D_{ang} \approx a_n \left(\frac{\pi}{2 n}+\varphi_T+\frac{3}{n} \arcsin\left(1+\frac{2 n}{\pi} \varphi_T\right) \right),
\label{eq:Dang}
\end{equation}
with
\begin{equation}
\varphi_T \approx -\frac{1}{n} \arcsin\left(\frac{4 \: \beta \: \rho_n \: a_S}{A \: S(a_n)}\right).
\label{eq:phiT}
\end{equation}
From equation \eqref{eq:phiT}, we deduce that there exists a critical value for the ratio $\frac{\beta}{A}$, i. e. $\left( \frac{\beta}{A} \right)_{crit}=\frac{S(a_n)}{4 \rho_n a_n}$. Whenever $\frac{\beta}{A} > \left( \frac{\beta}{A} \right)_{crit}$, the angular diameter vanishes, and there is no ovoid.\\
We have seen that the averaged system is invariant by a rotation of angle $\frac{2 \pi}{n}$. That is, two ovoids having the same radius $a_S$ have their centres S located $\frac{2 \pi}{n}$ apart along the line  $a=a_S$. Hence, to evaluate the probability for a given point of this line to be in the ovoid, we shall operate in the rectangular Van der Pol representation, and write that the capture probability $P_{capt, circle}$ is the ratio of the angular diameter of the ovoid to the distance between the centres of two contiguous ovoids. We get, when $\frac{\beta}{A} < \left( \frac{\beta}{A} \right)_{crit}$:
\begin{equation}
P_{capt, circle}=\frac{D_{ang}}{\frac{2 \pi}{n} a_n}=\frac{1}{4}+\frac{1}{2 \pi} \left(n \varphi_T +3 \arcsin \left(1+\frac{2 n}{\pi} \varphi_T \right)\right) ,
\label{eq:Pcapt_along_line}
\end{equation}
for initial condition  $(a=a_n, \; \varphi=\text{random value})$. \\
And for $\frac{\beta}{A} > \left( \frac{\beta}{A} \right)_{crit}$, we have $P_{capt}=0$.\\
Introduce a new parameter r, with $r=\frac{\frac{\beta}{A}}{\left( \frac{\beta}{A} \right)_{crit}}$. Hence $n \: \varphi_T \approx -\arcsin(r)$, and
\begin{equation}
P_{capt, circle}=\frac{1}{4}+\frac{1}{2 \pi} \left(-\arcsin(r)+3 \arcsin(1-\frac{2}{\pi} \arcsin(r))| \right).
\label{eq:Pcapt_along_line_r}
\end{equation}
\begin{figure}[h!]
\centering
\includegraphics[width=0.94\textwidth]{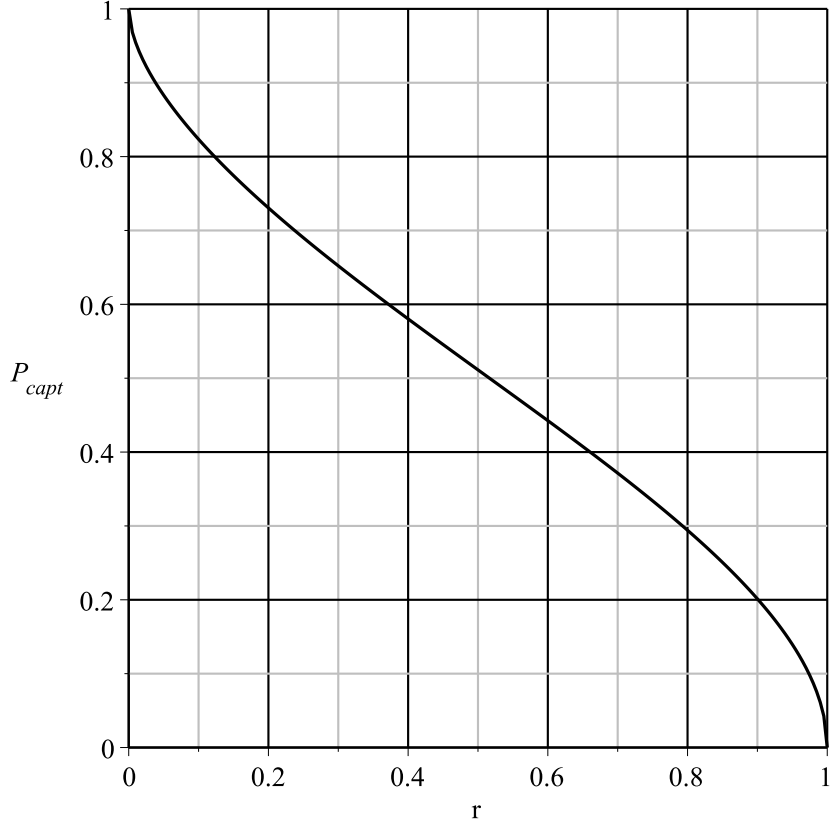}
\caption{Capture probability with initial condition $a=a_n$.}
\label{fig:Capt_probability_a_an}
\end{figure}
Figure \ref{fig:Capt_probability_a_an} shows $P_{capt}$ as a function of r.
\subsection{Capture probability with initial condition inside the annulus tangent to the ovoid.}
To assess this probability, although we operate in the polar Van der Pol representation, we shall carry out the calculus in the rectangular Van der Pol representation, to avoid a distortion due to the polar representation, where a given area next to the centre of coordinates is more probably hit than the same surface located far from the centre. We shall calculate the area of the ovoid, and form the ratio of this area to the area of the rectangle delimited by the left and right tangents to the ovoid (at points L and R in figure \ref{fig:Symbolic ovoid and Runge-Kutta solution}), and by the centres (points S in said figure) of two contiguous ovoids. \\
We define the radial diameter $D_{rad}$ of the ovoid as the distance between the left and right vertical tangents to the ovoid, i.e. approximately the distance LR. As the averaged system is invariant by a rotation of angle $\frac{2 \pi}{n}$, the distance between the centres of two contiguous ovoids is $\frac{2 \pi}{n} a_n$. The surface $S_{rect}$ of the rectangle is then $S_{rect}=\frac{2 \pi}{n} a_n D_{rad}$.\\
To calculate the area of the ovoid, we approximate it by an ellipse having as axes TB and LR of figure \ref{fig:Symbolic ovoid and Runge-Kutta solution}. We have $TB=D{ang}=a_n (\varphi_T-\varphi_B)$ and $LR=D{rad}=a_R-a_L$. The ovoid area is then $S_{ov}=\pi \frac{D_{ang}}{2} \frac{D_{rad}}{2}$.\\
Hence the capture probability with initial condition inside the annulus is $P_{capt,annulus1}=\frac{S_{ov}}{S_{rect}}=\frac{\pi \frac{D_{ang}}{2} \frac{D_{rad}}{2}}{\frac{2 \pi}{n} a_n D_{rad}}=\frac{n \: D_{ang}}{8 \: a_n}$. We can see that this expression does not depend on $D_{rad}$. This is because we approximated the ovoid by an ellipse, and took the annulus tangent to the ovoid as a delimiting area.\\
We see that, as the maximal possible value for $D_{ang}$ is $\frac{2 \pi}{n} a_n$, in which case two contiguous ovoids are in contact, the maximal possible value for $P_{capt,annulus1}$ is $\frac{n}{8 \: a_n} \frac{2 \pi}{n} a_n=\frac{\pi}{4}$.\\
Using the same parameter r than for the discussion of the probability with initial condition $a=a_n$, we get:
\begin{equation}
P_{capt,annulus1}=\frac{1}{8} \left(\frac{\pi}{2}-\arcsin(r)+3 \arcsin\left(1-\frac{2}{\pi} \arcsin(r) \right) \right),
\label{eq:Pcapt_annulus}
\end{equation}
with $r=\frac{\frac{\beta}{A}}{\left( \frac{\beta}{A} \right)_{crit}}$.\\
\begin{figure}[h!]
\centering
\includegraphics[width=0.94\textwidth]{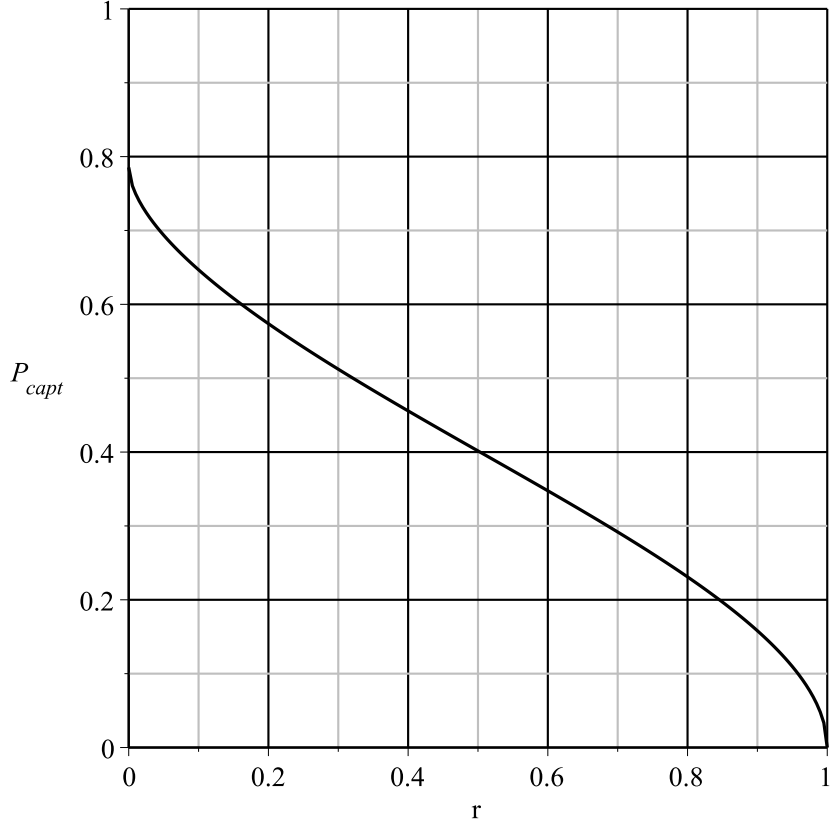}
\caption{Capture probability with initial condition inside the annulus.}
\label{fig:Capt_probability_annulus}
\end{figure}
Figure \ref{fig:Capt_probability_annulus} shows $P_{capt, annulus}$ as a function of r.
\subsection{Radial diameter.}
In this section, we shall discuss the radial diameter of the ovoid. This will enable us to assess, for a given value of n, the capture probability with initial condition inside the annulus delimiting an upper bound of the validity region of the averaging method, as represented in figure \ref{fig:multiple_ovoids_in_multiple_annuli}.\\
In figure \ref{fig:A_of_As_multiple_6_28_624}, we have seen that, for a given value of n that we choose to carry out the averaging method, we must constrain $a$ to be in a limited interval around $a_n$. An upper bound of this limiting interval is formed by the abscissae of $J_n$ and $J_{n+2}$.\\
To calculate the radial diameter of the ovoid, we shall write the equation giving $a_L$ and $a_R$. This will lead to a second-degree equation in $a$, whose solutions are $a_L$ and $a_R$. The difference of the roots of said equation is the radial diameter $D_{rad}=a_R-a_L$.\\
Assuming that the arguments of points L, S, and R are identical, i.e. $\varphi_L=\varphi_S=\varphi_R$, and that S(a) and D(a) are constants, respectively denoted S and D, in the ovoid region, we write that points T, L and R belong to the same integral curve:
\begin{equation}
\beta (\varphi_S-\varphi_T)-\frac{A S}{4 n \rho_n} \left( \frac{\cos(n \varphi_S)}{a}- \frac{\cos(n \varphi_T)}{a_T} \right)+\frac{3}{8} \frac{\mu}{\rho_n} \left(\frac{a^2}{2}-\frac{a_T^2}{2}-a_n^2 \left(\ln(a)-\ln(a_T) \right) \right)=0.
\label{eq:aL_aR}
\end{equation}
Besides, due to equations \eqref{eq:expression_of_phi_S_and_phi_T}, we have $\cos(n \varphi_S) \approx -\cos(n \varphi_T)=-\frac{a}{A D} \frac{3 \mu (a_T^2-a_n^2)}{2}$.\\
From equation \eqref{eq:giving_a_S_and_phi_S}, we have $sin(n \varphi_T) \approx -\frac{4 \rho_n \beta a_n}{A S(a_n)}$. Consequently, the coefficient $\frac{A S}{4 n \rho_n}$ from equation \eqref{eq:aL_aR} is approximated by $-\frac{1}{n} \: \frac{\beta a_n}{\sin(n \varphi_T)}$, and we can write, putting $\eta=\frac{\beta}{n} \cot(n \varphi_T)$ and $\zeta=\frac{3}{8} \frac{\mu}{\rho_n} a_n^2$:
\begin{equation}
\beta (\varphi_S-\varphi_T)-a_n \eta \left(\frac{1}{a_n}+\frac{1}{a_T} \right)+\frac{\zeta}{a_n^2} \left(\frac{a^2}{2}-\frac{a_T^2}{2}-a_n^2 (\ln(a)-\ln(a_T)) \right)=0.
\label{eq:aL_aR2}
\end{equation}
We shall study this equation in the neighbourhood of $a_n$, and we shall search for solutions in $a$ to the equation $y(a)=0$, denoting by $y(a)$ the left member of equation \eqref{eq:aL_aR2}.\\
Putting $a=a_n (1+\epsilon)$, we make a second-order Taylor development of $y(a)$ in the neighbourhood of $a_n$. We have: $y(a) \approx y(a_n)+\epsilon a_n y'(a_n)+\frac{(\epsilon a_n)^2}{2} y''(a_n)$. Replacing $\eta$ and $\zeta$ by their definition expressions and ordering with respect to $\epsilon$, we get:
\begin{equation}
(\zeta-\eta) \epsilon^2+\eta \epsilon-2 \eta=0.
\label{eq:epsil}
\end{equation}
The discriminant is $\Delta=\eta^2-4 (\zeta-\eta) (\beta (\varphi_S-\varphi_T)-2 \eta)$.
Let $\epsilon_L$ and $\epsilon_R$ denote the roots of this second-degree equation. We have, $a_R$ and $a_L$ denoting the values of $a$ at points R and L: $D_{rad}=a_R-a_L=a_n(1+\epsilon_R)-a_n(1+\epsilon_L)=a_n(\epsilon_R-\epsilon_L)=a_n \frac{\sqrt{\Delta}}{\zeta-\eta}$. Knowing that $\varphi_T-\varphi_S=\frac{\pi}{n}+2 \varphi_T$, we finally get:
\begin{equation}
D_{rad}=a_n \frac{\sqrt{\eta^2+4 (\zeta-\eta) (\beta \left(\frac{\pi}{n}+2 \varphi_T \right)+2 \eta)}}{\zeta-\eta},
\label{eq:Drad}
\end{equation}
with $\varphi_T \approx -\frac{1}{n} \arcsin(\frac{4 \beta \rho_n a_n}{A S(a_n)})$, $\eta=\frac{\beta}{n} \cot(n \varphi_T)$ and $\zeta=\frac{3}{8} \frac{\mu}{\rho_n} a_n^2$.

\subsection{Capture probability with initial condition inside the annulus delimiting an upper bound of the validity region of the averaging method.}
The total area of said annulus is $\pi (a_{n+2}^2-a_n^2)$. The total area of the n ovoids belonging to that annulus is $n S_{ov}=n \pi \frac{D_{ang}}{2} \frac{D_{rad}}{2}$. Hence, neglecting, inside the annulus, the area of the ovoids' upstream basins, the capture probability with initial condition in said annulus is $P_{capt,annulus2}=\frac{n}{4} \: \frac{D_{ang} D_{rad}}{a_{n+2}^2-a_n^2}$.\\
Replacing $a_n$ and $a_{n+2}$ by their developed expressions as given in equation \eqref{eq:definition_of_a_n}, we get:
\begin{equation}
P_{capt,annulus2}=-\frac{3 \mu}{64 \lambda^2} \frac{n^3 (n+2)^2} {n+1} D_{ang} D_{rad},
\end{equation}
with $\lambda=\frac{\nu}{\omega_0}$ and $D_{ang}$ given by equation \eqref{eq:Dang} and $D_{rad}$ given by equation \eqref{eq:Drad}.
\section{Future}
An interesting question is about systems with more than one minimum in the $A$-function, and therefore with more complicated implicit solutions and ovoid figures. These systems are important, because in them, the capture phenomenon can be initiated at surprisingly low amplitudes, and evolve into high-amplitude oscillations, taking advantage of the energy provided by the external harmonic excitation source. This kind of argumental oscillations is of interest to further delimit the hazardous parameter domains in civil engineering.
\bibliographystyle{plain}
\bibliography{Cintra_Argoul_debut_2017}

\begin{thebibliography}{10}

\bibitem{Bethenod_38_AS}
M.J. B\'ethenod.
\newblock Sur l'entretien du mouvement d'un pendule au moyen d'un courant
  alternatif de fr\'equence \'elev\'ee par rapport \`a sa fr\'equence propre.
\newblock {\em Comptes rendus hebdomadaires de l'Acad\'emie des sciences},
  207(19):847--849, November 1938.
\newblock (in French).

\bibitem{Bogolioubov_Book}
N.~Bogolioubov and I.~Mitropolski.
\newblock {\em Les me\'thodes asymptotiques en théorie des oscillations non
  linéaires}.
\newblock Gauthiers-Villars, 1962.

\bibitem{Cintra_Argoul_Dynolin_2013}
D.~Cintra and P.~Argoul.
\newblock Argumentary oscillation phenomenon.
\newblock In {\em (online publication)}, Lille, France, October 2013. Dynolin
  conference.

\bibitem{Cintra_Argoul_Vishno_2014}
D.~Cintra and P.~Argoul.
\newblock Argumentary {Duffing} oscillators - {Stable-regime} probability.
\newblock In {\em (online publication)}, Aix-en-Provence (France), June 2014.
  XIX th Vishno symposium.

\bibitem{Cintra_Argoul_Dynolin_2014}
D.~Cintra and P.~Argoul.
\newblock Six models of argumental oscillators - {Experimental} results.
\newblock In {\em (online publication)}, Paris, France, October 2014. Dynolin
  conference.

\bibitem{Cintra_Argoul_JVC_2016}
D.~Cintra and P.~Argoul.
\newblock Nonlinear argumental oscillators: A few examples of modulation via
  spatial position.
\newblock {\em Journal of Vibration and Control}, 2016.
\newblock (online publication, pre-printing).

\bibitem{Cornu}
A.~Cornu.
\newblock Sur la synchronisation des horloges de pr\'ecision et la distribution
  de l'heure.
\newblock {\em Journal de Physique Th\'eorique et Appliqu\'ee}, 7(1):231--239,
  1888.
\newblock (in French).

\bibitem{Cretin_Vernier}
B.~Cretin and D.~Vernier.
\newblock Quantized amplitudes in a nonlinear resonant electrical circuit.
\newblock In {\em 2009 Joint Meeting of the European Frequency and Time Forum
  and the IEEE International Frequency Control Symposium, vols 1 and 2}, volume
  1 \& 2, pages 797--800, Besan\c{c}on, France, April 2009. Joint Meeting of
  the 23rd European Frequency and Time Forum/IEEE International Frequency
  Control Symposium.

\bibitem{Soulier_25}
G.~Darrieus.
\newblock Joseph {B}\'ethenod - {Sa} vie, son oeuvre - {Pr\'esentation} \`a la
  {Soci\'et\'e francaise des Electriciens} le 2 d\'ecembre 1944.
\newblock {\em Technica}, 67:8, June-July 1945.
\newblock {The} presentation by A. Soulier is cited in this article about
  B\'ethenod (in French).

\bibitem{Doubochinski_Book}
D.~Doubochinski.
\newblock {\em Argumental oscillations. Macroscopic quantum effects.}
\newblock SciTech Library, August 2015.
\newblock To be translated in English.

\bibitem{Doubo_EDF_91}
D.B. Doubochinski and J.B. Doubochinski.
\newblock Amor\c{c}age argumentaire d'oscillations entretenues avec une s\'erie
  discr\`ete d'amplitudes stables.
\newblock {\em E.D.F. Bulletin de la direction des \'etudes et recherches,
  s\'erie C math\'ematiques, informatique}, 3:11--20, 1991.
\newblock (in French).

\bibitem{Fery}
Ch. F\'ery.
\newblock Sur quelques modes \'electriques d'entretien du pendule. {Pendule}
  sans lien mat\'eriel.
\newblock {\em Journal de Physique Th\'eorique et Appliqu\'ee}, 7(1):520--530,
  1908.
\newblock (in French).

\bibitem{Gradshteyn_07}
I.~M.~Ryzhik I.~S.~Gradshteyn.
\newblock {\em Table of Integrals, Series, and Products, Seventh Edition}.
\newblock Alan Jeffrey and Daniel Zwillinger, 2007.

\bibitem{Penner_Doubo_73}
D.~I. Penner, D.~B. Doubochinski, M.~I. Kozakov, A.~S. Vermel, and Yu.~V.
  Galkin.
\newblock Asynchronous excitation of undamped oscillations.
\newblock {\em Soviet Physics Uspekhi}, 16(1):158--160, July-August 1973.

\bibitem{Penner_Doubo_72}
D.~I. Penner, Ya.~B. Doubochinski, D.~B. Doubochinski, and M.~I. Kozakov.
\newblock Oscillations with self-regulating interaction time.
\newblock {\em Soviet Physics Doklady}, 17:541, December 1972.

\bibitem{Treilhou_00}
J.P. Treilhou, J.~Coutelier, J.J. Thocaven, and C.~Jacquez.
\newblock Payload motions detected by balloon-borne fluxgate-type
  magnetometers.
\newblock {\em Advances in Space Research}, 26(9):1423--1426, 2000.

\end{thebibliography}

\end{document}